\pdfoutput=1
\documentclass[12pt,a4paper]{article}

\usepackage{ifthen}
\newboolean{pdflatex}
\setboolean{pdflatex}{true}

\newboolean{articletitles}
\setboolean{articletitles}{true}

\newboolean{uprightparticles}
\setboolean{uprightparticles}{false}

\def\paperauthors{LHCb collaboration}
\def\paperasciititle{Measurement of tauL using the Bs -> J/psi eta decay mode}
\def\papertitle{Measurement of $\tau_\textrm{L}$ using the $\B_s^0 \rightarrow J/\psi \eta$ decay mode}
\def\paperkeywords{{High Energy Physics}, {LHCb}}
\def\papercopyright{\the\year\ CERN for the benefit of the LHCb collaboration}
\def\paperlicence{CC BY 4.0 licence}
\def\paperlicenceurl{https://creativecommons.org/licenses/by/4.0/}

\usepackage[top=1in, bottom=1.25in, left=1in, right=1in]{geometry}

\usepackage{microtype}
\usepackage{placeins}
\usepackage{lineno}
\usepackage{xspace}
\usepackage{caption}

\usepackage{graphicx}
\usepackage{color}
\usepackage{colortbl}

\usepackage{amsmath}
\usepackage{amssymb}
\usepackage{amsfonts}
\usepackage{upgreek}

\newcommand*\patchAmsMathEnvironmentForLineno[1]{
\expandafter\let\csname old#1\expandafter\endcsname\csname #1\endcsname
\expandafter\let\csname oldend#1\expandafter\endcsname\csname
end#1\endcsname
 \renewenvironment{#1}
   {\linenomath\csname old#1\endcsname}
   {\csname oldend#1\endcsname\endlinenomath}
}
\newcommand*\patchBothAmsMathEnvironmentsForLineno[1]{
  \patchAmsMathEnvironmentForLineno{#1}
  \patchAmsMathEnvironmentForLineno{#1*}
}
\AtBeginDocument{
\patchBothAmsMathEnvironmentsForLineno{equation}
\patchBothAmsMathEnvironmentsForLineno{align}
\patchBothAmsMathEnvironmentsForLineno{flalign}
\patchBothAmsMathEnvironmentsForLineno{alignat}
\patchBothAmsMathEnvironmentsForLineno{gather}
\patchBothAmsMathEnvironmentsForLineno{multline}
\patchBothAmsMathEnvironmentsForLineno{eqnarray}
}

\usepackage{hyperxmp}

\usepackage[pdftex,
            pdfauthor={\paperauthors},
            pdftitle={\paperasciititle},
            pdfkeywords={\paperkeywords},
            pdfcopyright={Copyright (C) \papercopyright},
            pdflicenseurl={\paperlicenceurl}]{hyperref}

\usepackage[colorinlistoftodos,textsize=scriptsize]{todonotes}

\usepackage[bottom,flushmargin,hang,multiple]{footmisc}

\usepackage[all]{hypcap}

\usepackage{xspace}
\usepackage{upgreek}

\def\lhcb   {\mbox{LHCb}\xspace}

\def\MagUp {\mbox{\em Mag\kern -0.05em Up}\xspace}

\ifthenelse{\boolean{uprightparticles}}
{

 \def\Pmu         {\ensuremath{\upmu}\xspace}

 \def\Ppi         {\ensuremath{\uppi}\xspace}

 \def\Pchi        {\ensuremath{\upchi}\xspace}
 \def\Ppsi        {\ensuremath{\uppsi}\xspace}

 \def\PDelta      {\ensuremath{\Delta}\xspace}
 \def\PXi         {\ensuremath{\Xi}\xspace}
 \def\PLambda     {\ensuremath{\Lambda}\xspace}
 \def\PSigma      {\ensuremath{\Sigma}\xspace}
 \def\POmega      {\ensuremath{\Omega}\xspace}
 \def\PUpsilon    {\ensuremath{\Upsilon}\xspace}
 \let\oldPi\Pi
 \def\PPi         {\ensuremath{\oldPi}\xspace}

 \def\PB      {\ensuremath{\mathrm{B}}\xspace}
 
 \def\PD      {\ensuremath{\mathrm{D}}\xspace}

 \def\PJ      {\ensuremath{\mathrm{J}}\xspace}
 \def\PK      {\ensuremath{\mathrm{K}}\xspace}

 \def\Pb      {\ensuremath{\mathrm{b}}\xspace}
 \def\Pc      {\ensuremath{\mathrm{c}}\xspace}

 \def\Pi      {\ensuremath{\mathrm{i}}\xspace}

 \def\Ps      {\ensuremath{\mathrm{s}}\xspace}

 \def\thebaroffset{0.0em}
}
{

 \def\Pmu         {\ensuremath{\mu}\xspace}

 \def\Ppi         {\ensuremath{\pi}\xspace}

 \def\Pchi        {\ensuremath{\chi}\xspace}
 \def\Ppsi        {\ensuremath{\psi}\xspace}
 
 \mathchardef\PDelta="7101
 \mathchardef\PXi="7104
 \mathchardef\PLambda="7103
 \mathchardef\PSigma="7106
 \mathchardef\POmega="710A
 \mathchardef\PUpsilon="7107
 \mathchardef\PPi="7105
 
 \def\PB      {\ensuremath{B}\xspace}
 
 \def\PD      {\ensuremath{D}\xspace}

 \def\PJ      {\ensuremath{J}\xspace}
 \def\PK      {\ensuremath{K}\xspace}

 \def\Pb      {\ensuremath{b}\xspace}
 \def\Pc      {\ensuremath{c}\xspace}

 \def\Pi      {\ensuremath{i}\xspace}

 \def\Ps      {\ensuremath{s}\xspace}

 \def\thebaroffset{0.18em}
}
\newcommand{\offsetoverline}[2][\thebaroffset]{\kern #1\overline{\kern -#1 #2}}

\makeatletter
\ifcase \@ptsize \relax
  \newcommand{\miniscule}{\@setfontsize\miniscule{4}{5}}
\or
  \newcommand{\miniscule}{\@setfontsize\miniscule{5}{6}}
\or
  \newcommand{\miniscule}{\@setfontsize\miniscule{5}{6}}
\fi
\makeatother

\DeclareRobustCommand{\optbar}[1]{\shortstack{{\miniscule (\rule[.5ex]{1.25em}{.18mm})}
  \\ [-.7ex] $#1$}}

\def\mup        {{\ensuremath{\Pmu^+}}\xspace}
\def\mun        {{\ensuremath{\Pmu^-}}\xspace}

\def\squark    {{\ensuremath{\Ps}}\xspace}

\def\cquark    {{\ensuremath{\Pc}}\xspace}

\def\bquark    {{\ensuremath{\Pb}}\xspace}

\def\pion   {{\ensuremath{\Ppi}}\xspace}

\def\pip    {{\ensuremath{\pion^+}}\xspace}
\def\pim    {{\ensuremath{\pion^-}}\xspace}

\def\kaon    {{\ensuremath{\PK}}\xspace}

\def\KorKbar {\kern \thebaroffset\optbar{\kern -\thebaroffset \PK}{}\xspace}

\def\Kp      {{\ensuremath{\kaon^+}}\xspace}
\def\Km      {{\ensuremath{\kaon^-}}\xspace}

\def\D       {{\ensuremath{\PD}}\xspace}

\def\DorDbar {\kern \thebaroffset\optbar{\kern -\thebaroffset \PD}\xspace}

\def\Dp      {{\ensuremath{\D^+}}\xspace}
\def\Dm      {{\ensuremath{\D^-}}\xspace}

\def\DpDm    {\ensuremath{\Dp {\kern -0.16em \Dm}}\xspace}

\def\Dsp     {{\ensuremath{\D^+_\squark}}\xspace}
\def\Dsm     {{\ensuremath{\D^-_\squark}}\xspace}

\def\B       {{\ensuremath{\PB}}\xspace}
\def\Bbar    {{\ensuremath{\offsetoverline{\PB}}}\xspace}

\def\BorBbar {\kern \thebaroffset\optbar{\kern -\thebaroffset \PB}\xspace}
\def\Bz      {{\ensuremath{\B^0}}\xspace}

\def\Bd      {{\ensuremath{\B^0}}\xspace}

\def\BdorBdbar {\kern \thebaroffset\optbar{\kern -\thebaroffset \Bd}\xspace}
\def\Bu      {{\ensuremath{\B^+}}\xspace}

\def\Bp      {{\ensuremath{\Bu}}\xspace}

\def\Bs      {{\ensuremath{\B^0_\squark}}\xspace}
\def\Bsb     {{\ensuremath{\Bbar{}^0_\squark}}\xspace}
\def\BsorBsbar {\kern \thebaroffset\optbar{\kern -\thebaroffset \Bs}\xspace}
\def\Bc      {{\ensuremath{\B_\cquark^+}}\xspace}

\def\jpsi     {{\ensuremath{{\PJ\mskip -3mu/\mskip -2mu\Ppsi}}}\xspace}

\def\chic     {{\ensuremath{\Pchi_\cquark}}\xspace}

\def\chicone  {{\ensuremath{\Pchi_{\cquark 1}}}\xspace}
\def\chictwo  {{\ensuremath{\Pchi_{\cquark 2}}}\xspace}

\def\Upsilonres  {{\ensuremath{\PUpsilon}}\xspace}
\def\Y#1S{\ensuremath{\PUpsilon{(#1S)}}\xspace}

\def\Lz          {{\ensuremath{\PLambda}}\xspace}

\def\LorLbar     {\kern \thebaroffset\optbar{\kern -\thebaroffset \PLambda}\xspace}

\def\Lb           {{\ensuremath{\Lz^0_\bquark}}\xspace}

\newcommand{\decay}[2]{\ensuremath{#1\!\to #2}\xspace}

\def\to                 {\ensuremath{\rightarrow}\xspace}

\def\CP                {{\ensuremath{C\!P}}\xspace}

\def\AT#1     {\ensuremath{A_{\mathrm{T}}^{#1}}\xspace}

\def\C#1      {\ensuremath{\mathcal{C}_{#1}}\xspace}
\def\Cp#1     {\ensuremath{\mathcal{C}_{#1}^{'}}\xspace}
\def\Ceff#1   {\ensuremath{\mathcal{C}_{#1}^{\mathrm{(eff)}}}\xspace}
\def\Cpeff#1  {\ensuremath{\mathcal{C}_{#1}^{'\mathrm{(eff)}}}\xspace}
\def\Ope#1    {\ensuremath{\mathcal{O}_{#1}}\xspace}
\def\Opep#1   {\ensuremath{\mathcal{O}_{#1}^{'}}\xspace}

\newcommand{\aunit}[1]{\ensuremath{\text{\,#1}}}

\newcommand{\tev}{\aunit{Te\kern -0.1em V}\xspace}
\newcommand{\gev}{\aunit{Ge\kern -0.1em V}\xspace}
\newcommand{\mev}{\aunit{Me\kern -0.1em V}\xspace}
\newcommand{\kev}{\aunit{ke\kern -0.1em V}\xspace}
\newcommand{\ev}{\aunit{e\kern -0.1em V}\xspace}

\newcommand{\mevc}{\ensuremath{\aunit{Me\kern -0.1em V\!/}c}\xspace}
\newcommand{\gevc}{\ensuremath{\aunit{Ge\kern -0.1em V\!/}c}\xspace}
\newcommand{\mevcc}{\ensuremath{\aunit{Me\kern -0.1em V\!/}c^2}\xspace}
\newcommand{\gevcc}{\ensuremath{\aunit{Ge\kern -0.1em V\!/}c^2}\xspace}

\def\fb   {\ensuremath{\aunit{fb}}\xspace}
\def\invfb   {\ensuremath{\fb^{-1}}\xspace}

\def\ps   {\ensuremath{\aunit{ps}}\xspace}
\def\fs   {\aunit{fs}}

\newcommand{\stat}{\aunit{(stat)}\xspace}
\newcommand{\syst}{\aunit{(syst)}\xspace}

\newcommand{\chisq}{\ensuremath{\chi^2}\xspace}

\newcommand{\chisqip}{\ensuremath{\chi^2_{\text{IP}}}\xspace}

\def\gsim{{~\raise.15em\hbox{$>$}\kern-.85em
          \lower.35em\hbox{$\sim$}~}\xspace}
\def\lsim{{~\raise.15em\hbox{$<$}\kern-.85em
          \lower.35em\hbox{$\sim$}~}\xspace}

\def\sPlot{\mbox{\em sPlot}\xspace}

\def\pt         {\ensuremath{p_{\mathrm{T}}}\xspace}

\def\ptot       {\ensuremath{p}\xspace}

\def\evtgen     {\mbox{\textsc{EvtGen}}\xspace}

\def\geant      {\mbox{\textsc{Geant4}}\xspace}

\def\photos     {\mbox{\textsc{Photos}}\xspace}

\def\pythia     {\mbox{\textsc{Pythia}}\xspace}

\def\tell1  {TELL1\xspace}
\def\ukl1   {UKL1\xspace}

\newcommand{\lhcborcid}[1]{\href{https://orcid.org/#1}{\hspace*{0.1em}\raisebox{-0.45ex}{\includegraphics[width=1em]{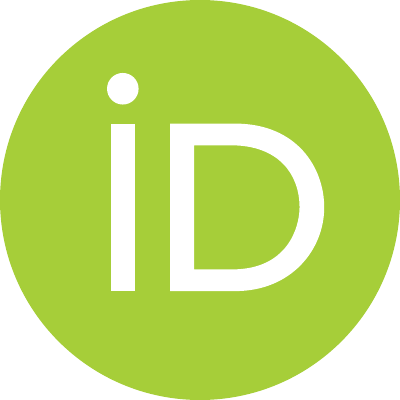}}}}

\usepackage{cite}
\usepackage{mciteplus}

\begin{document}

\renewcommand{\thefootnote}{\fnsymbol{footnote}}
\setcounter{footnote}{1}

\begin{titlepage}
\pagenumbering{roman}

\vspace*{-1.5cm}
\centerline{\large EUROPEAN ORGANIZATION FOR NUCLEAR RESEARCH (CERN)}
\vspace*{1.5cm}
\noindent
\begin{tabular*}{\linewidth}{lc@{\extracolsep{\fill}}r@{\extracolsep{0pt}}}
\vspace*{-1.5cm}\mbox{\!\!\!\includegraphics[width=.14\textwidth]{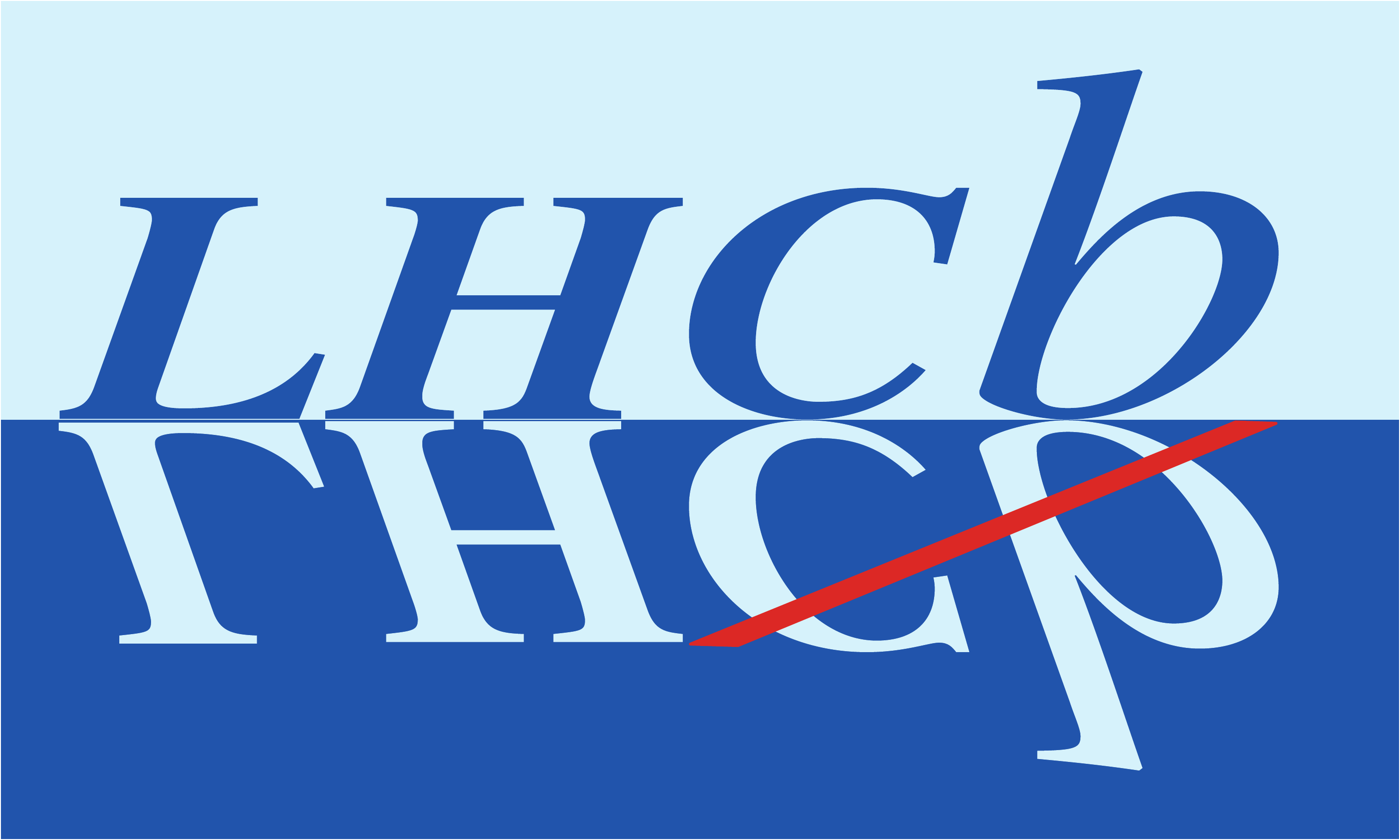}} & &
\\
 & & CERN-EP-2022-099 \\
 & & LHCb-PAPER-2022-010 \\
 & & \today \\
\end{tabular*}

\vspace*{4.0cm}

{\normalfont\bfseries\boldmath\huge
\begin{center}
  \papertitle
\end{center}
}

\vspace*{2.0cm}
\begin{center}
%In the footnote, replace 'paper' by 'Letter' in case of submission to PRL or PLB 
% Edit macro in main.tex to keep metadata correct
\paperauthors\footnote{This paper is dedicated to our friend and colleague Marie-No\"{e}lle Minard.} \footnote{Authors are listed at the end of the paper.}
\end{center}

\vspace{\fill}

\begin{abstract}
  \noindent
    Using a proton-proton collision data sample collected by the \lhcb detector and corresponding to an integrated luminosity of $5.7~\textrm{fb}^{-1}$, the lifetime of the light $\Bs$ mass eigenstate, $\tau_\textrm{L}$, is measured using the $B^0_s \rightarrow J/\psi\eta$ decay mode to be
\begin{displaymath}
\tau_{\textrm{L}} = 1.445 \pm 0.016 \textrm{(stat)} \pm 0.008 \textrm{(syst)} \ps. \nonumber
\end{displaymath}
A combination of this result with a previous \lhcb analysis using an independent dataset corresponding to
3 fb$^{-1}$ of integrated  luminosity gives
\begin{displaymath}
\tau_{\textrm{L}} = 1.452 \pm 0.014 \pm 0.007 \pm 0.002 \ps, \nonumber
\end{displaymath}
where the first uncertainty is statistical, the second due to the uncorrelated part of the systematic uncertainty and the third due to the correlated part of the systematic uncertainty.
\end{abstract}

\vspace*{2.0cm}

\begin{center}
  Published in Eur.~Phys.~J.~C83~(2023)~629
\end{center}

\vspace{\fill}

{\footnotesize
\centerline{\copyright~\papercopyright. \href{\paperlicenceurl}{\paperlicence}.}}
\vspace*{2mm}

\end{titlepage}

\newpage
\setcounter{page}{2}
\mbox{~}

\renewcommand{\thefootnote}{\arabic{footnote}}
\setcounter{footnote}{0}

\cleardoublepage

\pagestyle{plain}
\setcounter{page}{1}
\pagenumbering{arabic}

\section{Introduction}
\label{sec:Introduction}
In the Standard Model (SM), the $\Bs$ and $\Bsb$ flavour eigenstates can be expressed as a linear combination of the heavy ({\rm H}) and light ({\rm L}) mass eigenstates with decay widths $\Gamma_{\rm H}$ and $\Gamma_{\rm L}$, respectively. A sizeable difference between these decay widths is predicted \cite{Lenz:2019lvd}. The effective lifetime of a $\Bs$ meson in a specific decay mode is measured by fitting the decay time distribution with a single exponential function \cite{Fleischer:2011cw,
  Fleischer:2011ib}. The experimental data on the weak mixing phase, $\phi_s$, are consistent with the SM prediction that \CP violation in $\Bs - \Bsb$ mixing is small \cite{Lenz:2019lvd}. Consequently, the mass eigenstates are also \CP eigenstates to better than per mille level and the effective lifetime measured in $B_s^0 \rightarrow J/\psi \eta$ decays is equal, to good approximation, to $\tau_{\rm L} = 1/\Gamma_{\rm L}$. Using the SM prediction for $\Delta \Gamma_s$ from Ref.~\cite{Lenz:2019lvd}, and the value of  $\tau_{\Bs}/\tau_{\Bd}$ given in Ref.~\cite{Kirk:2017juj}, together with the measured lifetime of the \Bd meson~\cite{PDG2020} gives $\tau_{\rm L} = 1.422 \pm 0.013 \ps$.
Measurements of $\tau_{\rm L}$ have been reported by the LHCb collaboration using the tree-level decay modes  $\Bs \rightarrow \jpsi \eta$ \cite{LHCb-PAPER-2016-017}, $\Bs\rightarrow\Dsp \Dsm$ \cite{LHCb-PAPER-2013-060} and the loop dominated
$\Bs \rightarrow K^+ K^-$ \cite{LHCb-PAPER-2014-011} where hadronic uncertainties are more important. These measurements use data collected during Run 1 of the LHC at centre-of-mass energies, $\sqrt{s} = 7 \,$ and $8 \,$TeV, corresponding to an integrated luminosity of $3\invfb$. The lifetime of the heavy mass eigenstate, $\tau_{\rm H}$, has been measured using the $\Bs \rightarrow \jpsi f_0(980)$ mode \cite{LHCb-PAPER-2012-017,D0:2016nbv, CMS:2017ygm}. Improving the precision of $\tau_{\rm L}$ and $\tau_{\rm H}$ enables more stringent tests of the consistency between direct measurements of the decay-width difference, $\Delta \Gamma_s = \Gamma_{\rm L} - \Gamma_{\rm H}$, in $\Bs \rightarrow \jpsi \phi$ decays and those inferred from effective lifetimes.

In this paper, $\tau_{\rm L}$ is measured in the $\Bs \rightarrow \jpsi \eta$ decay mode using a data sample,  corresponding to an integrated luminosity of $5.7 \invfb$, collected in proton-proton ($pp$) collisions at $\sqrt{s} = 13 \, \textrm{TeV}$ during Run 2 of the LHC (2015--2018). The analysis builds on the Run 1 study described in Ref.~\cite{LHCb-PAPER-2016-017} which used around $3 \invfb$ collected at $\sqrt{s} = 7,8 \, \textrm{TeV}$. The $\jpsi$ meson is reconstructed via the dimuon decay mode and the $\eta$ meson through its decay to a pair of photons.
\section{Detector and simulation}
\label{sec:detector}
The \lhcb detector~\cite{LHCb-DP-2012-002,LHCb-DP-2014-002} is a single-arm forward spectrometer covering the \mbox{pseudorapidity} range $2<\eta <5$, designed for the study of particles containing \bquark or \cquark quarks. It includes a high-precision tracking system consisting of a silicon-strip vertex detector (VELO) surrounding the $pp$ interaction region, a large-area silicon-strip detector (TT) located upstream of a dipole magnet with a bending power of approximately $4{\mathrm{\,Tm}}$, and three stations of silicon-strip detectors and straw drift tubes placed downstream of the magnet. The tracking system provides a measurement of the momentum, \ptot, of charged particles with a relative uncertainty that varies from 0.5\,\% at low momentum to 1.0\,\% at 200\gevc. Large samples of  $\jpsi\to\mup\mun$ and $B^+ \to \jpsi K^+$ decays, collected concurrently with the data set used here, were used to calibrate the momentum scale of the spectrometer using the procedure discussed in Ref.~\cite{LHCb-PAPER-2013-011}. The relative accuracy of this procedure is determined using samples of other fully reconstructed $b$-hadrons and \Upsilonres mesons, and is estimated to be $3 \times 10^{-4}$. For $\bquark$-hadron decay modes such $B^+ \to \jpsi K^+$ the mass resolution agrees between data and simulation to better than $10 \%$.

Various charged hadrons are distinguished using information from two ring-imaging Cherenkov detectors. In addition, photons, electrons, and hadrons are identified by a calorimeter system consisting of scintillating-pad and preshower detectors, an electromagnetic and a hadronic calorimeter. The calorimeter response is calibrated
using samples of $\pi^0 \rightarrow \gamma \gamma$ decays \cite{LHCb-DP-2020-001}. For this analysis, a further calibration is made using an inclusive sample of $\eta \rightarrow \gamma\gamma$ decays, which results in a precision of $0.1\,\%$ on the neutral energy scale. The mass resolution on the $\eta$ meson found in simulation (data) is $22.9 \mevcc$ ($23.9 \mevcc$).  Muons are identified by a system composed of alternating layers of iron and multiwire proportional chambers.

 The online event selection is performed by a trigger, which consists of a hardware stage followed by a two-level software stage \cite{LHCb-DP-2019-001}. An alignment and calibration of the detector is performed in near real-time with the results used in the software trigger~\cite{Borghi:2017hfp}. The same alignment and calibration information is propagated to the offline reconstruction, ensuring consistent information between the trigger and offline software. In this analysis, candidate events are required to pass the hardware trigger, which selects muon and dimuon candidates with high transverse momentum, $\pt$, using information from the muon system. The first stage of the software trigger performs a partial event reconstruction and requires events to have two well-identified oppositely charged muons with an invariant mass larger than $2.7 \gevcc$. The second stage performs a
full event reconstruction.  Events are retained for further
processing if they contain a displaced $\jpsi \rightarrow \mu^+ \mu^-$
candidate. The decay vertex is required to be well separated
from each reconstructed primary vertex (PV) of the proton-proton
interaction by requiring the distance between the PV and the $\jpsi$
decay vertex  divided by its
uncertainty (refered to as the decay-length significance or DLS) to be greater than three. This introduces a non-uniform
efficiency for $\bquark$-hadron candidates that have a decay time less than
$\sim 0.4 \ps$.

Simulated $pp$ collisions are generated using
\pythia~\cite{Sjostrand:2006za,*Sjostrand:2007gs}  with a specific
\lhcb configuration~\cite{LHCb-PROC-2010-056}.  Decays of hadronic
particles are described by \evtgen~\cite{Lange:2001uf}, in which
final-state radiation is generated using
\photos~\cite{Golonka:2005pn}. The interaction of the generated
particles with the detector, and its response, are implemented using the \geant toolkit~\cite{Allison:2006ve,
  *Agostinelli:2002hh} as described in Ref.~\cite{LHCb-PROC-2011-006}. Other sources of background, such as those from inclusive $b \rightarrow \chi_c$ transitions,
where the $\chic$ decays radiatively to a $\jpsi$ meson, are studied using the RapidSim fast simulation package \cite{Cowan:2016tnm}.

\section{Selection}
\label{sec:selection}
As in the LHCb Run 1 analysis of this mode \cite{LHCb-PAPER-2016-017}, a two-step procedure is used to optimize the selection of $\Bs \rightarrow
\jpsi \eta$ decay candidates. These studies use simulated signal samples together with the
high-mass sideband of the data ($5650 < m(\jpsi \eta) < 5850 \mevcc$),
which is not used in the subsequent determination of $\tau_{\rm L}$. In
the first step, loose selection criteria are applied that
reduce background significantly whilst retaining
high signal efficiency. Subsequently, a
multivariate analysis (MVA) is used to reduce further the combinatorial
background. This is optimised using pseudoexperiments to obtain the best precision on the measured value of $\tau_{\rm L}$. Compared to the Run 1 analysis, the modifications to both steps improve the signal efficiency and background rejection.

The selection starts with a pair of oppositely charged particles,
identified as muons, that form a common decay vertex. To ensure a high
efficiency the muon candidates are required to have a pseudorapidity between
2.0 and 4.6. The invariant mass of the dimuon candidate must be within $\pm 50 \mevcc$ of the known $\jpsi$ mass
\cite{PDG2020}.

Photons are selected from well-identified neutral clusters reconstructed in the electromagnetic
calorimeter~\cite{LHCb-DP-2014-002} that have a transverse energy in excess of $300 \mev$. Candidate $\eta\to\gamma \gamma$ decays are selected from
diphoton combinations with an invariant mass within $70\mevcc$ of the known $\eta$~mass \cite{PDG2020} and with a transverse momentum larger than $2.1 \gevc$.

The $\jpsi$ and $\eta$ candidates are combined to form candidate $\Bs$ mesons, which are required to have $\pt$ larger than $2.5 \gevc$. The $\Bs$ candidate is assigned to the PV with the smallest $\chisqip$, where \chisqip\ is defined as the difference in the vertex-fit \chisq to a given PV reconstructed with and without the candidate being considered. Though lifetime biasing, a loose requirement of $\chisqip<25$ is applied since it is effective in removing combinatorial background. In addition, to ensure the $\Bs$ candidate is matched to the correct PV, if there is another PV for  which  $\chisqip<50$ in the event, the candidate is rejected. A kinematic fit is performed to improve the invariant mass resolution~\cite{Hulsbergen:2005pu}, where the intermediate resonance masses are constrained to their known values. The $\chi^2$~per degree of freedom of this fit is required to be less than five. The measured decay time of the candidate must be between $0.4 \ps$ and $10 \ps$. These requirements define a time range where the acceptance is reasonably uniform.

The second step of the selection process is based on a Multilayer Perceptron neural network \cite{Hocker:2007ht}, which is trained using the simulated signal sample and the high-mass sideband of the data for background. Fourteen variables, chosen as they give good separation between signal and background, are used as input to the neural network. It has been checked that these variables are well modelled by the simulation using the $\Bs \rightarrow \jpsi \eta$ signal after the preselection and subtracting the background with the \sPlot \cite{Pivk:2004ty} technique and also with the $\Bu \rightarrow \jpsi K^{+}$ control channel.

These are chosen as they give good separation between signal and background, and are known to be well modelled by the simulation. The variables include information on the candidate kinematics, particle identification, vertexing and track quality. An important change compared to the Run 1 analysis is the addition of information related to the isolation of the $b$-hadron candidate. This is found to be highly effective at suppressing combinatorial background and compensates for the higher multiplicity in the Run 2 environment.

The requirement on the MVA output was chosen to minimize the
expected statistical uncertainty on the fitted value of  $\tau_{\textrm{L}}$ using pseudoexperiments. The chosen value removes over $99 \, \%$ of background candidates whilst retaining over $80 \, \%$
of simulated signal decays.  After applying these requirements $5 \, \%$ of events contain multiple candidates from which only one, chosen at random, is kept.

\section{Fit model}
\label{sec:fitmodel}
\begin{figure}[htbp!]
\begin{center}
\vspace{3mm}
\includegraphics[width=0.48\textwidth]{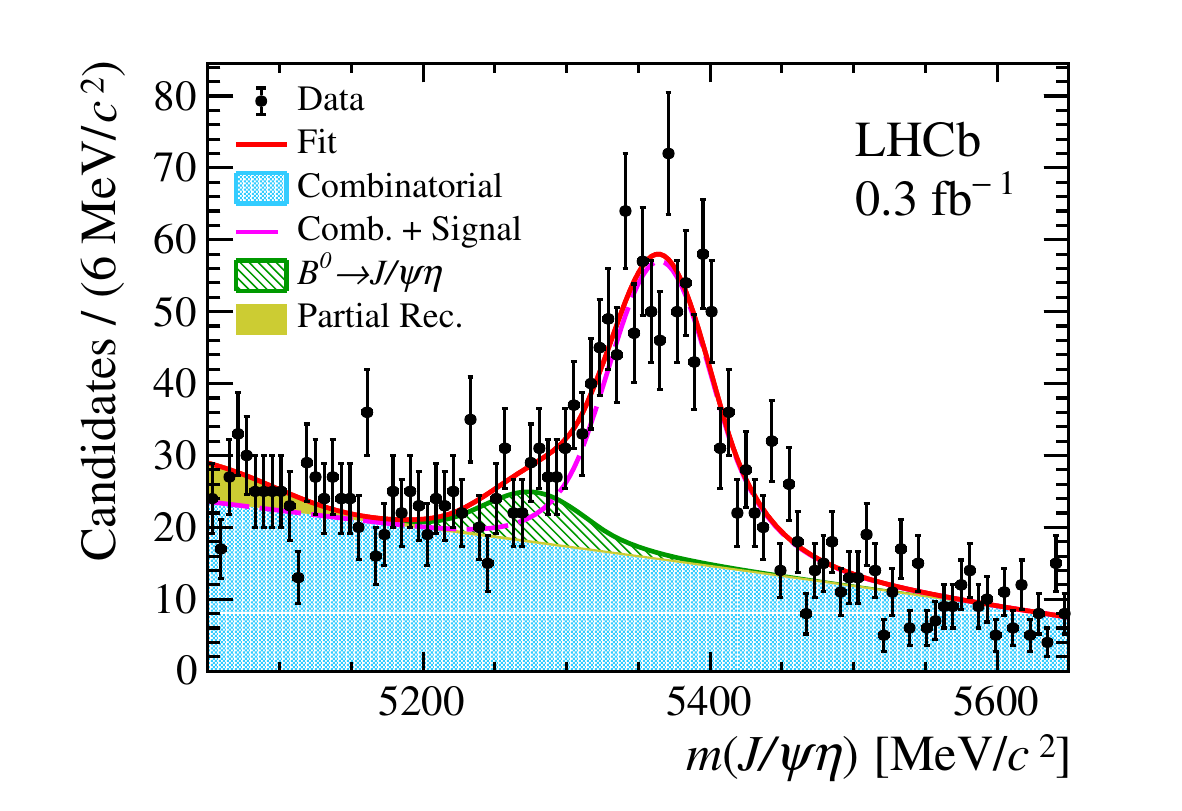}
\includegraphics[width=0.48\textwidth]{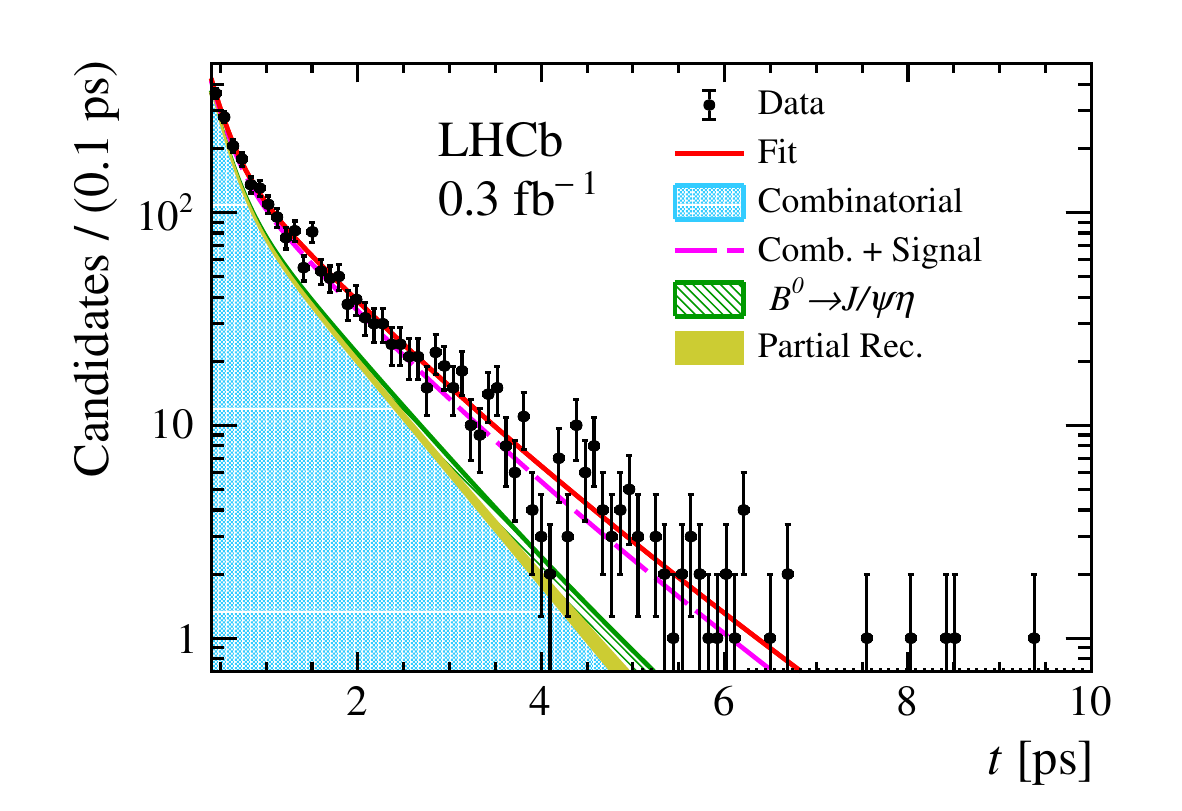}
\includegraphics[width=0.48\textwidth]{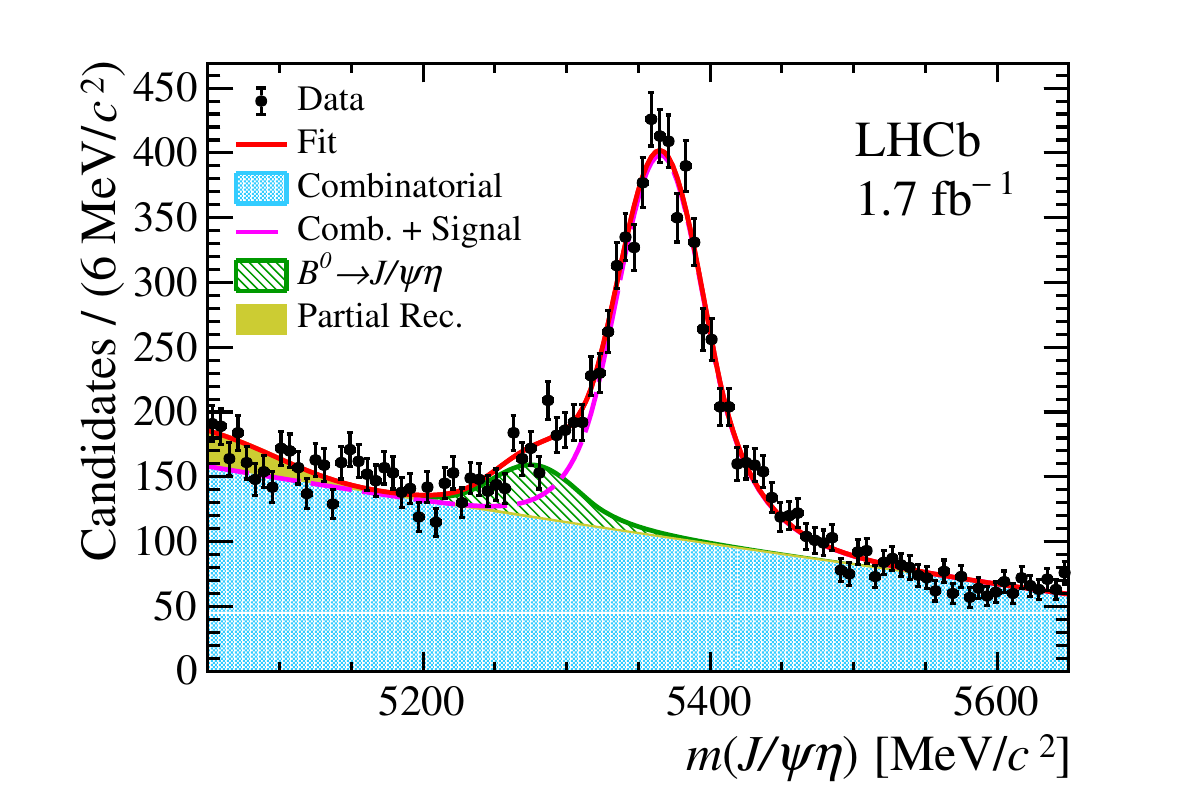}
\includegraphics[width=0.48\textwidth]{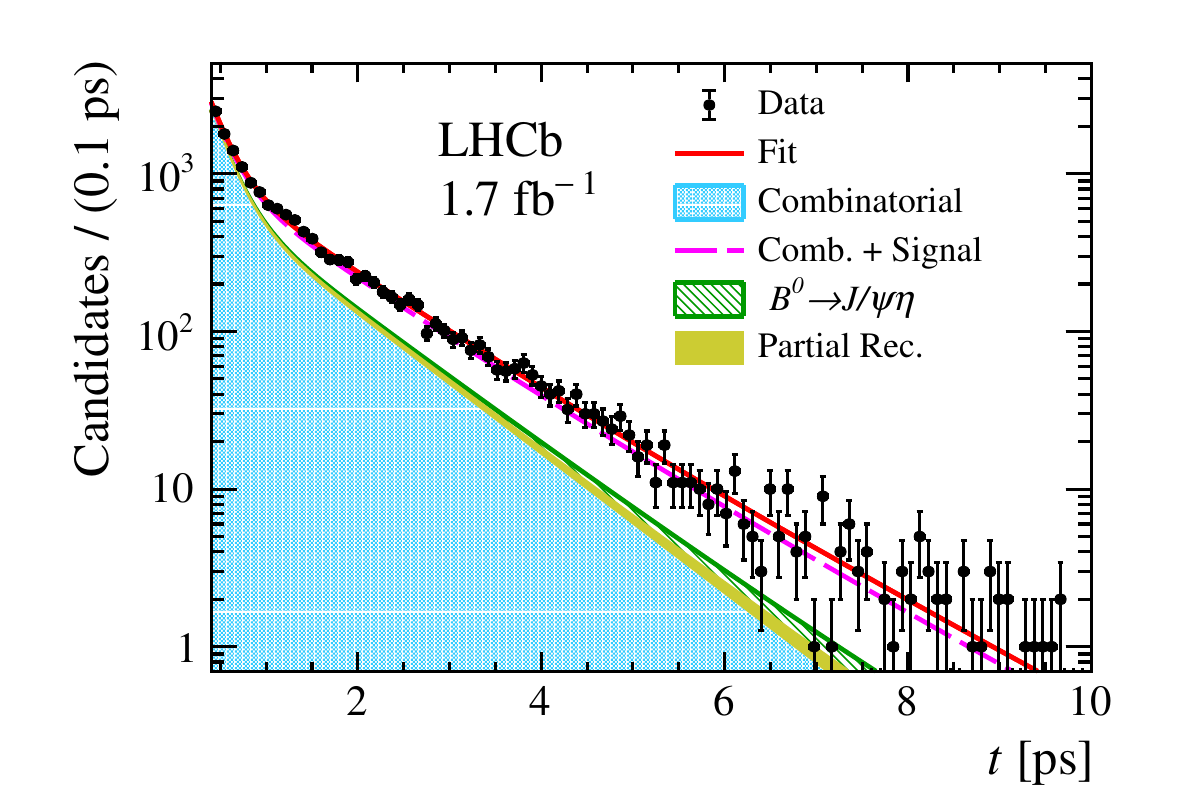}
\includegraphics[width=0.48\textwidth]{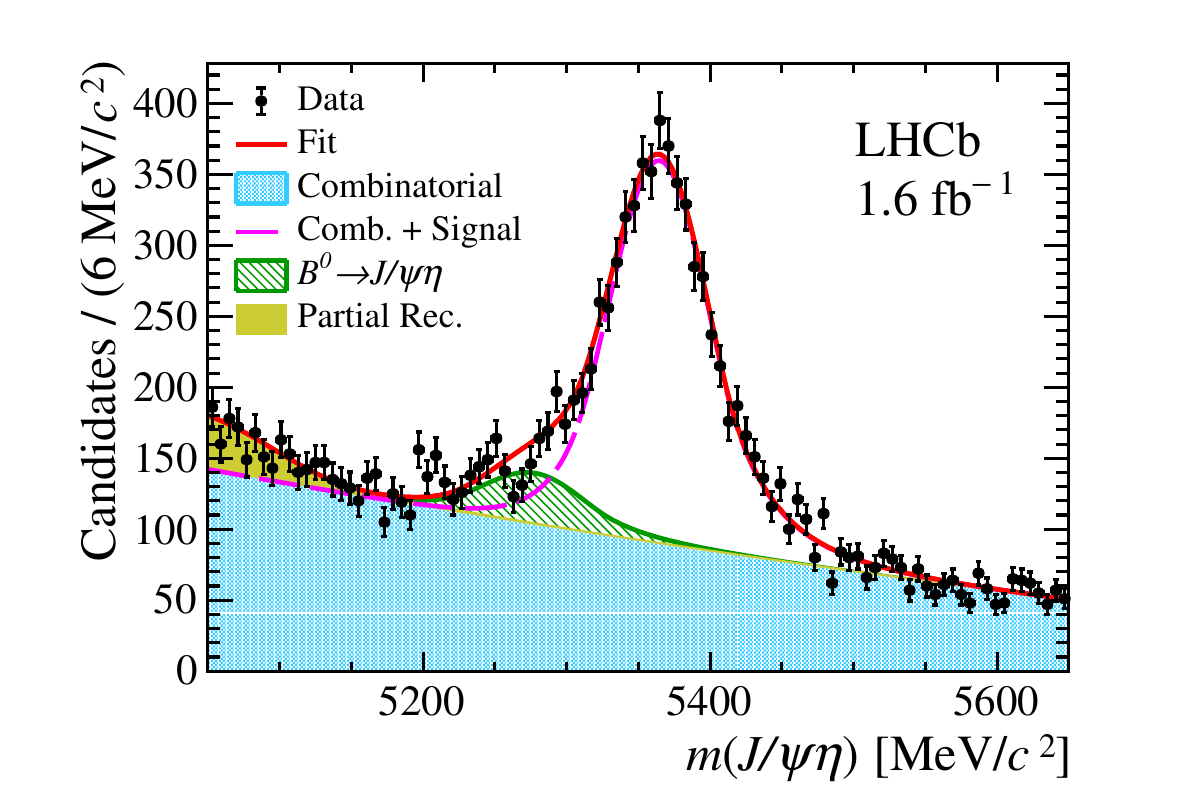}
\includegraphics[width=0.48\textwidth]{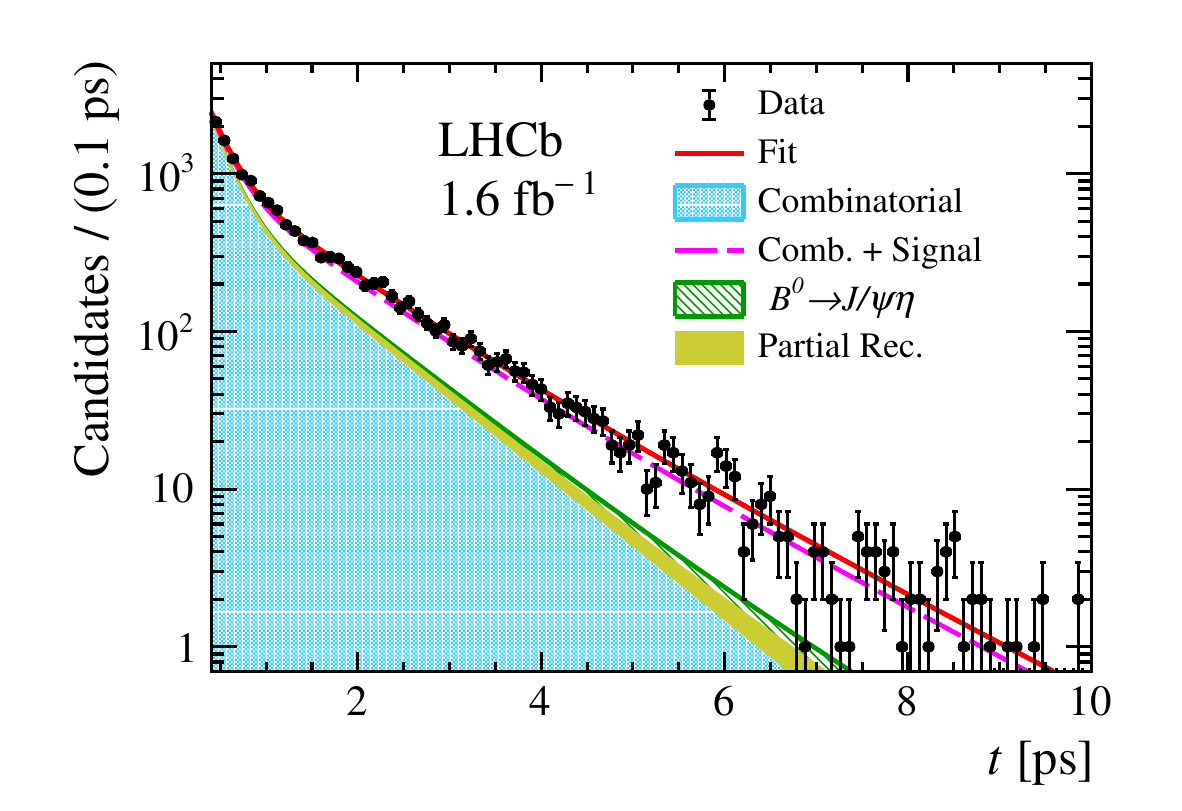}
\includegraphics[width=0.48\textwidth]{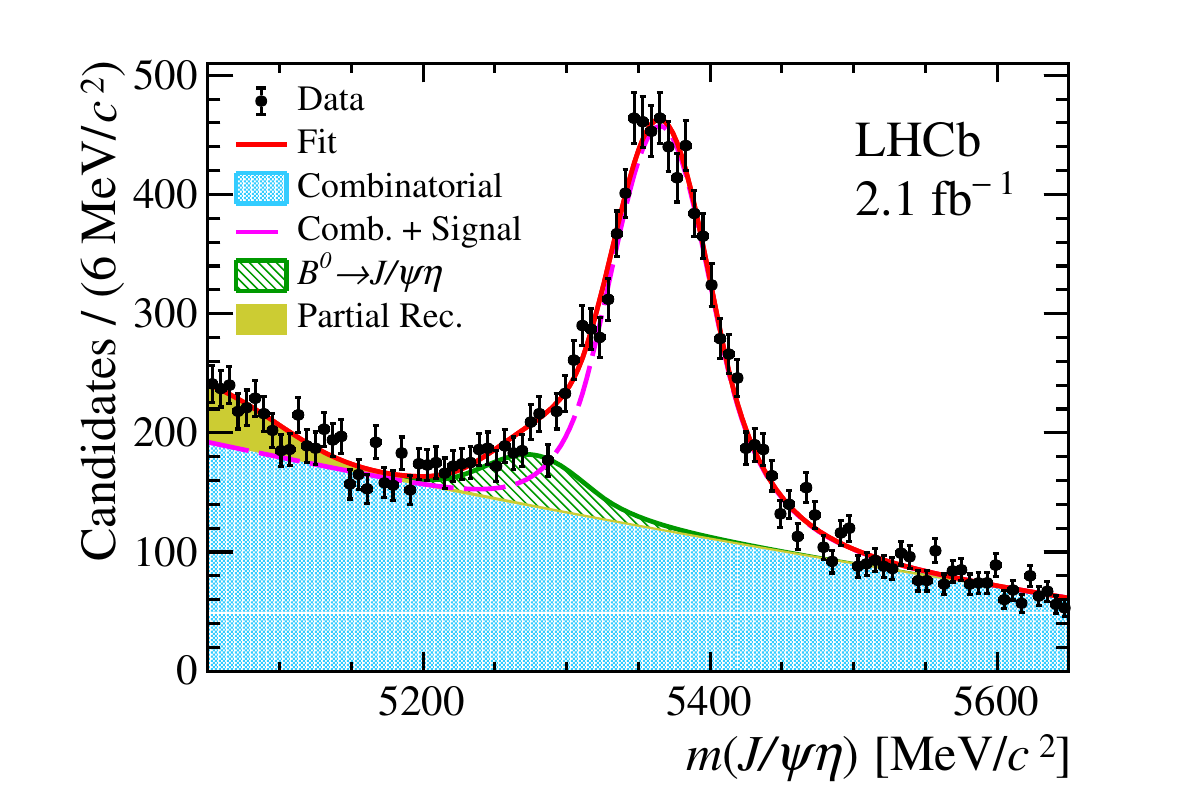}
\includegraphics[width=0.48\textwidth]{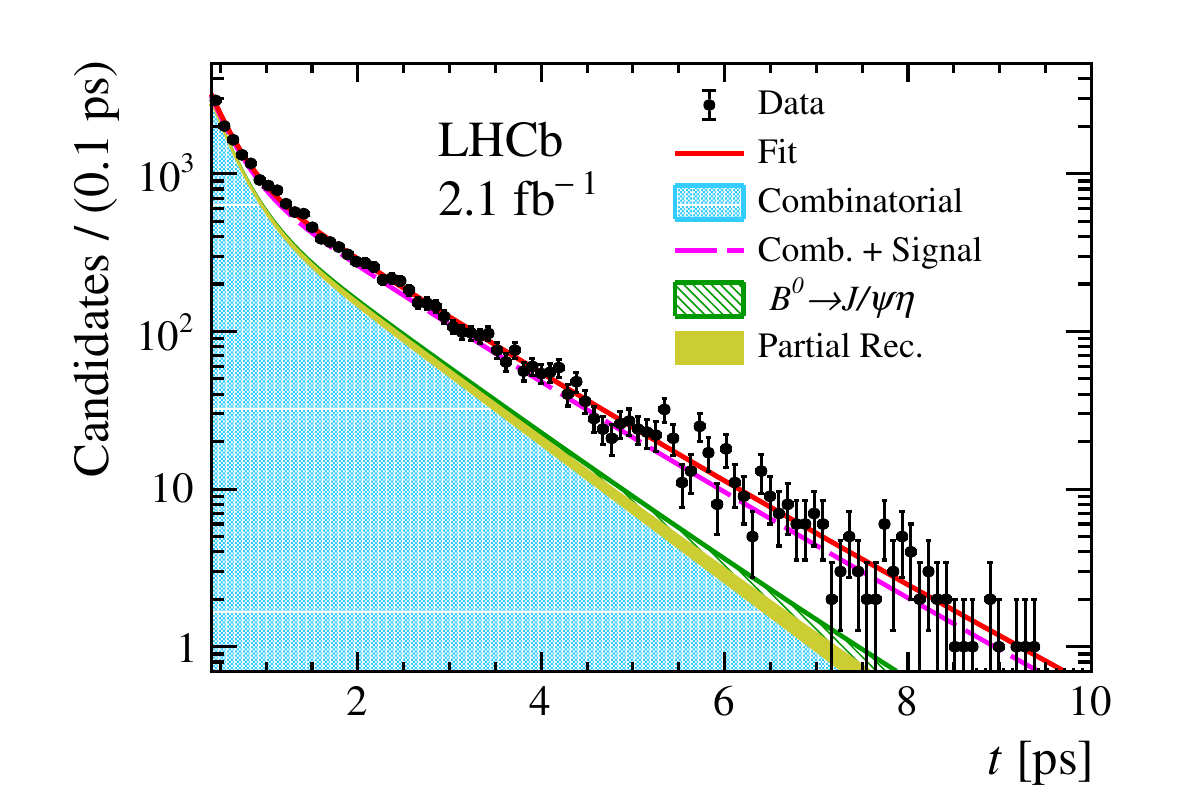}
\vspace{-2mm}
\caption{Distributions of (left) invariant mass and (right) decay time by year: (top row) 2015, (second row) 2016, (third row) 2017 and (bottom row) 2018. }
\label{fig:yearFits}
\end{center}
\end{figure}
Figure \ref{fig:yearFits} shows the $\jpsi\eta$ invariant mass and decay time distributions along with the
fit projections for the four years of running (2015, 2016, 2017 and 2018). The value of $\tau_{\rm L}$ is determined from a two-dimensional unbinned maximum likelihood fit to the distributions of the $\B^0_{(s)}$ candidate invariant mass, $m$ and decay time. To allow for variations in running conditions, the dataset is divided into the four years of running which are fitted simultaneously. The fit is performed for candidates with $ 5050 <m(\jpsi
\eta) < 5650 \mevcc$ and $0.4 < t < 10 \ps $.
The fit model has five components: the $\Bs \rightarrow \jpsi \eta$ signal, the $\Bz \rightarrow \jpsi \eta$ decay, partially reconstructed
$\Bs \rightarrow \chi_{c1,c2} \eta$ decays with the subsequent decay $\chi_c \rightarrow \jpsi \gamma$, partially reconstructed $\Bs \rightarrow \jpsi \phi$ with $\phi \rightarrow \eta \gamma$ and combinatorial background.

To account for the detector resolution, the decay time distribution of each component is convolved with a Gaussian function with a width of $52 \fs$ determined using the simulation. The variation of the detector efficiency as a function of decay time for each year of running is accounted for by an acceptance function, $A_{\textrm{tot}}$. This is expressed as the product of four components. Firstly, the
reconstruction efficiency of the vertex detector is known to decrease
as the distance of closest approach of the decay products to the
$pp$ beam-line increases \cite{LHCb-PAPER-2013-065, LHCb-PAPER-2016-017}. This effect is parameterised with a second order polynomial. Improvements in the track reconstruction used in Run 2 led to a reduction for this effect by a factor of three compared to the Run 1 analysis \cite{LHCb-DP-2019-001}. It is further reduced by the choice of the pseudorapidity range for the muons and cross-checked with the $\Bu \rightarrow \jpsi K^+$ decay mode. Secondly, the requirement on the DLS applied in the trigger leads to inefficiency at small proper times. Its effect is minimised by the requirement of $t > 0.4 \ps$. The residual effect of this requirement, $A_{\rm DLS}$, is modelled using the simulation and calibrated in data using a large sample of $\Bu \rightarrow \jpsi K^+$ decays. Furthermore, the requirement on the $\Bs$ candidate $\chisqip$ leads to an efficiency, $A_{\textrm{IP} \chi^2}$, that decreases linearly at higher time. This component is parameterised using simulated signal decays. Finally, the inclusion of the $\Bs$ candidate $\chisqip$ in the MVA leads to a linear acceptance correction ($A_{\rm MVA}$) which is determined using the simulation.

\begin{figure}[htb!]
\begin{center}
\resizebox{5.5in}{!}{\includegraphics{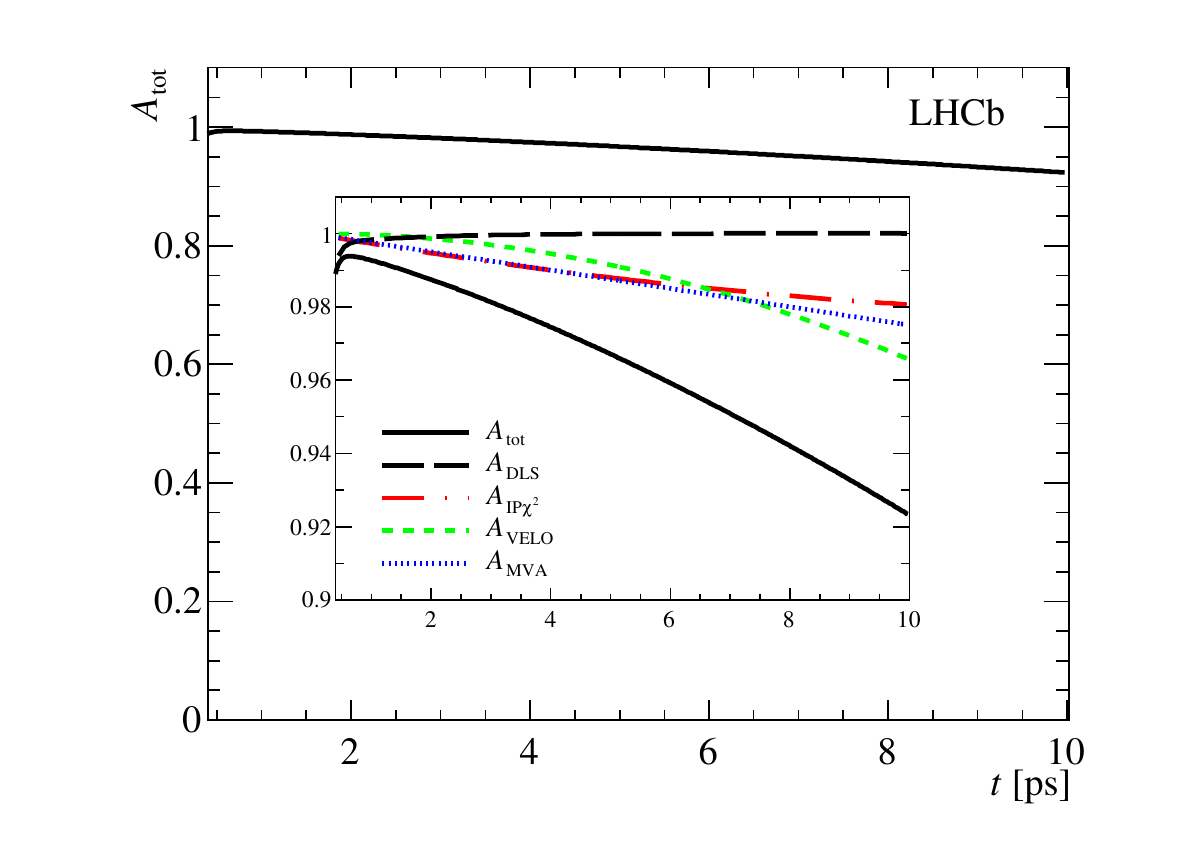}}
\vspace{-5mm}
\caption{\small Total acceptance function, $A_{\textrm{tot}}$ for the  2016 data taking period. The insert shows the four individual components of the acceptance that are multiplied to give $A_{\textrm{tot}}$. The acceptance functions for the other years are similar. }
\label{fig:accdata}
\end{center}
\end{figure}

Figure~\ref{fig:accdata} shows the overall acceptance curve, $A_{\rm tot} = A_{\textrm{VELO}}\cdot A_{\textrm{DLS}} \cdot A_{\textrm{IP} \chi^2} \cdot A_{\textrm{MVA}} $ obtained for the 2016 dataset. If the data were not corrected for the acceptance, then the bias on $\tau_{\rm L}$, evaluated by fitting the simulated data with and without the correction, would be $18 \fs$. Different, but qualitatively similar, decay time acceptance curves are evaluated for each year and used in the decay time fit.

The invariant mass distribution for the $\Bs \rightarrow \jpsi \eta$
signal is parameterised by a Double Sided Crystal Ball (DSCB) function. This is a generalization of the Crystal Ball function with power law tails on both sides of the mass peak \cite{Skwarnicki:1986xj}. Alternative parametrisations including the Bukin function~\cite{Bukin} and a skewed Student's t-distribution \cite{jbes.2009.06161,mnsc.44.12.1650} are considered as systematic variations.
 In the fit to the data, the tail parameters of the DSCB
distribution are fixed to the values obtained from simulation, while the mean and width parameter are freely varied in the fit.  The decay time distribution for the signal component is modelled with an exponential function
convolved with the detector resolution and multiplied by the detector acceptance, as discussed above.

The second component in the fit accounts for the $\Bz \rightarrow \jpsi \eta$ decay. Since the invariant mass resolution is approximately $36 \mevcc$, this component partially overlaps with the $\Bs$ signal mode.
The same DSCB mass shape as for the $\Bs$ signal decay is used to model this contribution, with a constraint on the mass resolution between the modes, as obtained from simulation. In the fit, the difference between the positions of the $\Bs$ and $\Bz$ mass peaks is Gaussian-constrained to the known value $M(\Bs) - M(\Bd) =  87.22 \pm 0.16 \mevcc$~\cite{PDG2020}. The decay time of the $\Bz$ component is modelled with an exponential convolved with the same acceptance function and detector resolution as for the $\Bs$ component. The lifetime is Gaussian-constrained to the known value, $\tau(\Bz) = 1.519 \pm 0.005 \ps$~\cite{PDG2020}. Finally, the yield of this component is parameterised as the product of the $\Bs$ yield for each year multiplied by a common factor between the years, $f_r$, which is left free. The shape of the combinatorial background in mass and time is studied using the high-mass sideband and mixing samples of $\jpsi$ and $\eta$ mesons selected in data. Based upon these studies this component is modelled by a second-order Chebyshev polynomial in mass and the sum of two exponentials in decay time.

The RapidSim \cite{Cowan:2016tnm} package is used to study backgrounds from partially reconstructed $b$-hadron decays and two sources are identified. The first is the decay $\Bs \rightarrow \jpsi \phi$ with the subsequent decay $\phi \rightarrow \eta \gamma$. Due to the missing photon, and the mass difference between the $\phi$ and $\eta$ mesons, this background has a maximum invariant mass of $5200 \, \mevcc$. The shape of this component in mass is modelled with a bifurcated Gaussian function. Its lifetime parameter is Gaussian-constrained to the measured value of the lifetime in the $\Bs \rightarrow \jpsi \phi$ channel, $\tau = 1.480 \pm 0.007 \ps$~\cite{HFLAV18} multiplied by a factor of 0.99, determined from the simulation, that accounts for the missing photon energy. The relative yield of this component to the signal is Gaussian-constrained to be $f_\phi = (0.6 \pm 0.1) \%$ using the known branching fractions \cite{PDG2020} and the relative efficiency from the simulation. The second source of partially reconstructed background is due to $\Bs \rightarrow \chi_{c1,c2} \eta$ decays. The branching fractions for these decay modes are not measured but the rate for the decay to the $\chicone$ meson is most likely higher than that of the $\chictwo$ decay mode from spin suppression arguments. Consequently, the yield of the $\chicone$ component, $N_{\chicone}$, is left to freely vary in the fit. The yield of the $\chictwo$ component is Gaussian-constrained to be $f_{\chictwo} N_{\chicone}$ where $f_{\chictwo} = 0.1 \pm 0.1$ based on the values seen for other decay modes \cite{PDG2020}. The shape of these components in mass is modelled by an error function, based on simulation. The unknown lifetime parameter of this component is left free in the fit. The validity of these assumptions is tested as part of the study of systematic uncertainties.

In total the simultaneous fit to the four years of running has 40 free parameters. The correctness of the fit procedure is validated using the full simulation and pseudoexperiments. No significant bias compared to the uncertainty of the measurement is found, and the uncertainties estimated by the fit are found to be accurate.
\section{Results and systematic uncertainties}
\label{sec:results}
\begin{table}[b!]
\caption{\small Systematic uncertainties on the lifetime
  measurement in $\fs$. Uncertainties less than $0.1\fs$ are indicated with a dash.}
\label{tab:systematics}
\begin{center}
\small
\begin{tabular}{l|c}
Source &  Uncertainty [fs] \\
\hline
Simulated sample sizes &  5.2 \\
$A_{\rm VELO}$   & 1.1 \\
$A_{\rm DLS}$ & -- \\
$A_{\rm IP\chisq}$  & 0.4 \\
$A_{\rm MVA}$ &  1.7 \\
$\Bu$ lifetime & 4.0 \\
Time resolution model & 0.3 \\
VELO half alignment & 3.8 \\
$\tau$ for $\Bs \rightarrow \chi_c \eta$ component & 0.7  \\
Mass model & 0.8 \\
$\Bd$ component & 0.4 \\
Momentum scale   &  --  \\
$z$-scale         & 0.3 \\
Data-simulation \chisqip differences & 0.1 \\
Mass-time correlation & 0.5
\\
$\Bc$ component & 1.0 \\
\hline
Quadrature sum & 8.0 \\
\end{tabular}
\end{center}
\end{table}

The fit gives a measured lifetime of $\tau_{\rm L} = 1.445 \pm 0.016\ps$. The values of other fitted parameters, such as the mass resolution are in good agreement with the expectation from the simulation. As discussed in Section~\ref{sec:fitmodel}, uncertainties on physics parameters are propagated to the statistical uncertainty via Gaussian constraints. The dominant uncertainty arises from the size of the simulation sample used to model the acceptance function. The uncertainty due to each of the individual acceptance corrections is estimated by resampling the covariance matrix of te determined PDF and by changing the functional form used for each component. The check of acceptance shape due to the VELO tracking efficiency using the $\Bu \rightarrow \jpsi K^+$ data is limited by the knowledge of the $\Bu$ lifetime giving a further $4 \fs$ uncertainty.

Possible biases from misalignment of the VELO detector halves are evaluated by dividing the data into three categories according to which side of the VELO the tracks traverse and re-running the fit. This gives a systematic uncertainty of $3.8 \fs$. In the fit to the decay time the correlation between the time resolution and its uncertainty is ignored. The possible bias from this assumption is evaluated to be less than $0.3 \fs$. Varying the time resolution, in the range $40 - 60 \fs$, does not change the result and no uncertainty is assigned.

Uncertainties arising from the modelling of the signal and
background mass distributions are evaluated using the discrete profiling method described in Ref.~\cite{Dauncey:2014xga}. In the default fit the yield of the $\Bd \rightarrow \jpsi \eta$ component is left free. Constraining it using the knowledge of the branching fractions in Ref.~\cite{PDG2020} and the fragmentation fractions given in Ref.~\cite{LHCb-PAPER-2020-046} changes $\tau_{\rm L}$ by $0.4 \fs$ which is assigned as a systematic.

Further small uncertainties arise from the limited knowledge of the length scale of the detector along the beam axis ($z$-scale), the momentum scale for charged particles and
the neutral particle energy scale. An additional $1 \fs$ systematic uncertainty arises from the effect of $\Bs$ mesons produced via $\Bc$ decays \cite{LHCb-PAPER-2019-013}.

\begin{figure}[t!]
\begin{center}
{\includegraphics[width=0.48\textwidth]{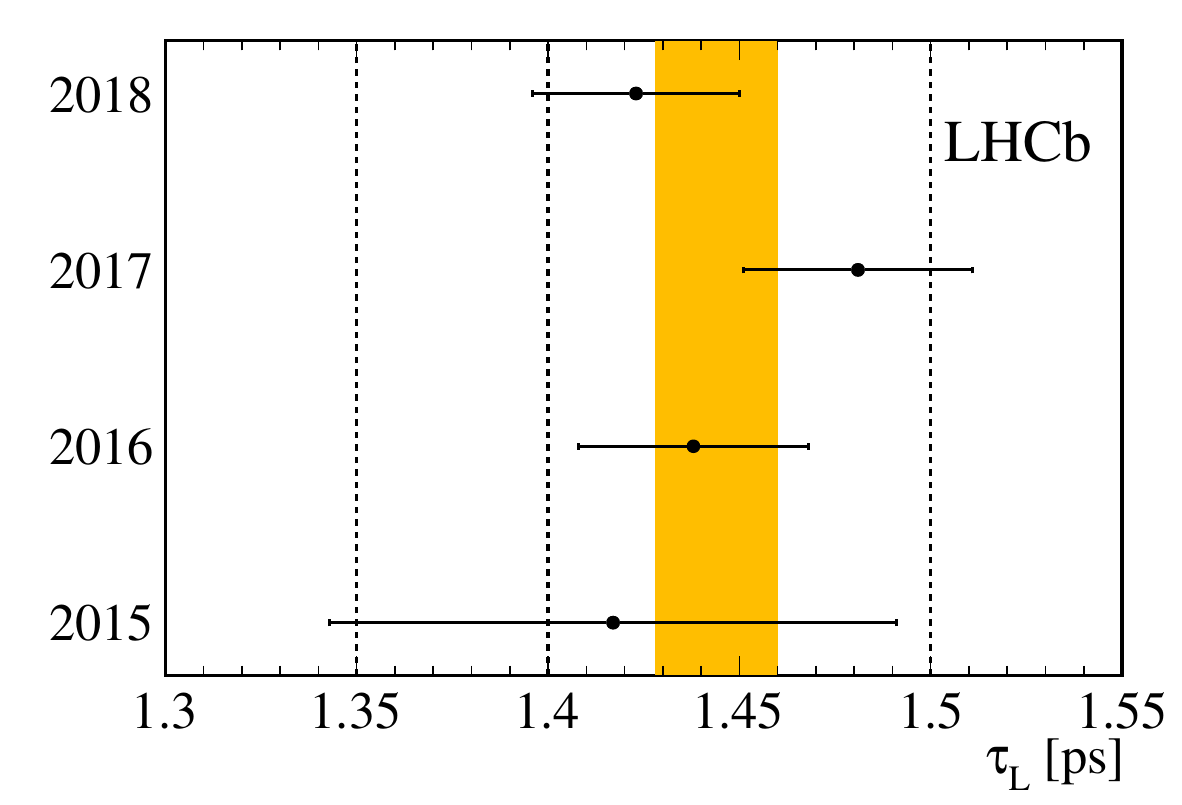}}
{\includegraphics[width=0.48\textwidth]{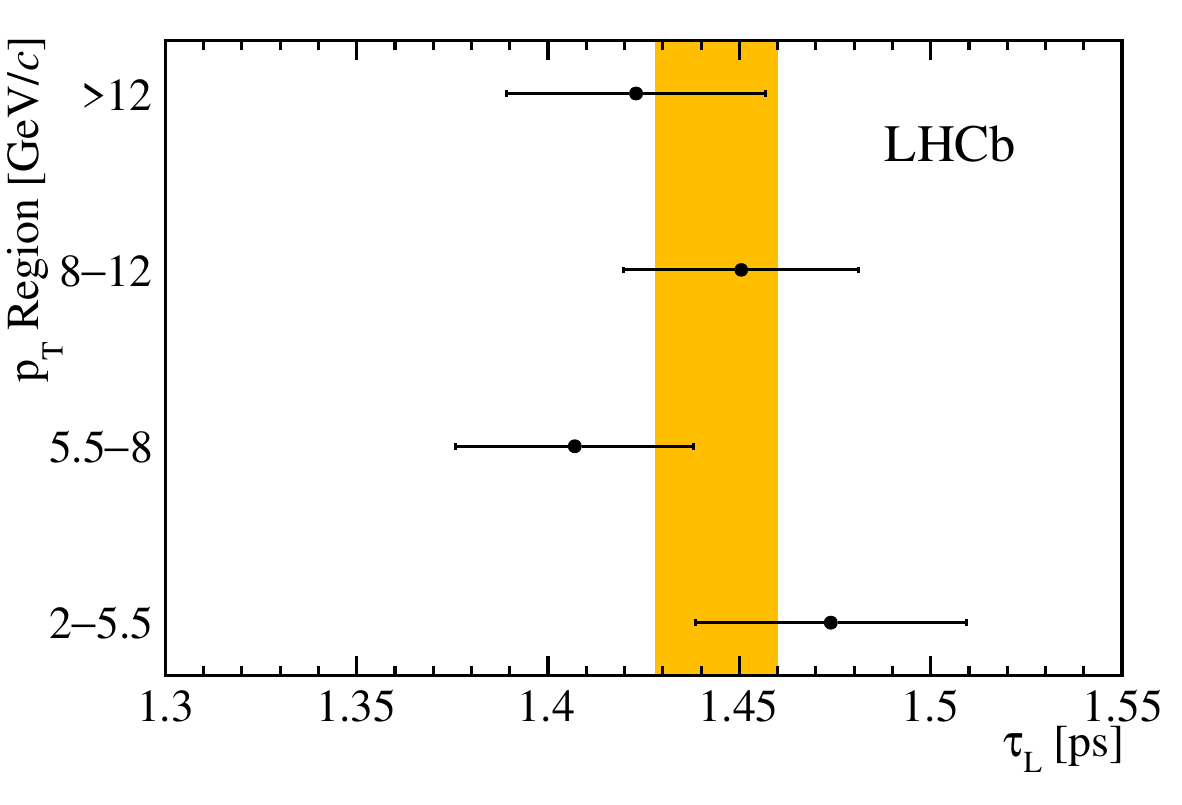}}
\vspace{-3mm}
\caption{\small Values of $\tau_{\rm L}$ obtained by dividing the data according to (left) the four years of data-taking and (right) regions in \Bs \pt. In both plots the statistical uncertainty for each point is shown. The $1 \sigma$ error bands of the statistical uncertainty on the combined results are shown in yellow. The $\chi^2$ probability of the measurements is $51 \, \%$($49 \, \%$) for fits to the data divided by year (\Bs \pt).}
\label{fig:tauyear}
\end{center}
\end{figure}

The stability of the result is tested by dividing the data according to the year of running (Fig.~\ref{fig:tauyear}, left), the $\pt$ of the $\Bs$ candidate (Fig.~\ref{fig:tauyear}, right) and by varying the requirements on the MVA, the $\chisqip$ and the value of the minimum decay time requirement. No significant change in the final result is found and hence no further systematic uncertainty is assigned.

The systematic uncertainties are summarized in Table~\ref{tab:systematics}. Adding them in quadrature leads to a total systematic uncertainty of $8.0 \fs$.

\section{Summary}
\label{sec:summary}
\begin{figure}[htb!]
\begin{center}
\resizebox{5.0in}{!}{\includegraphics{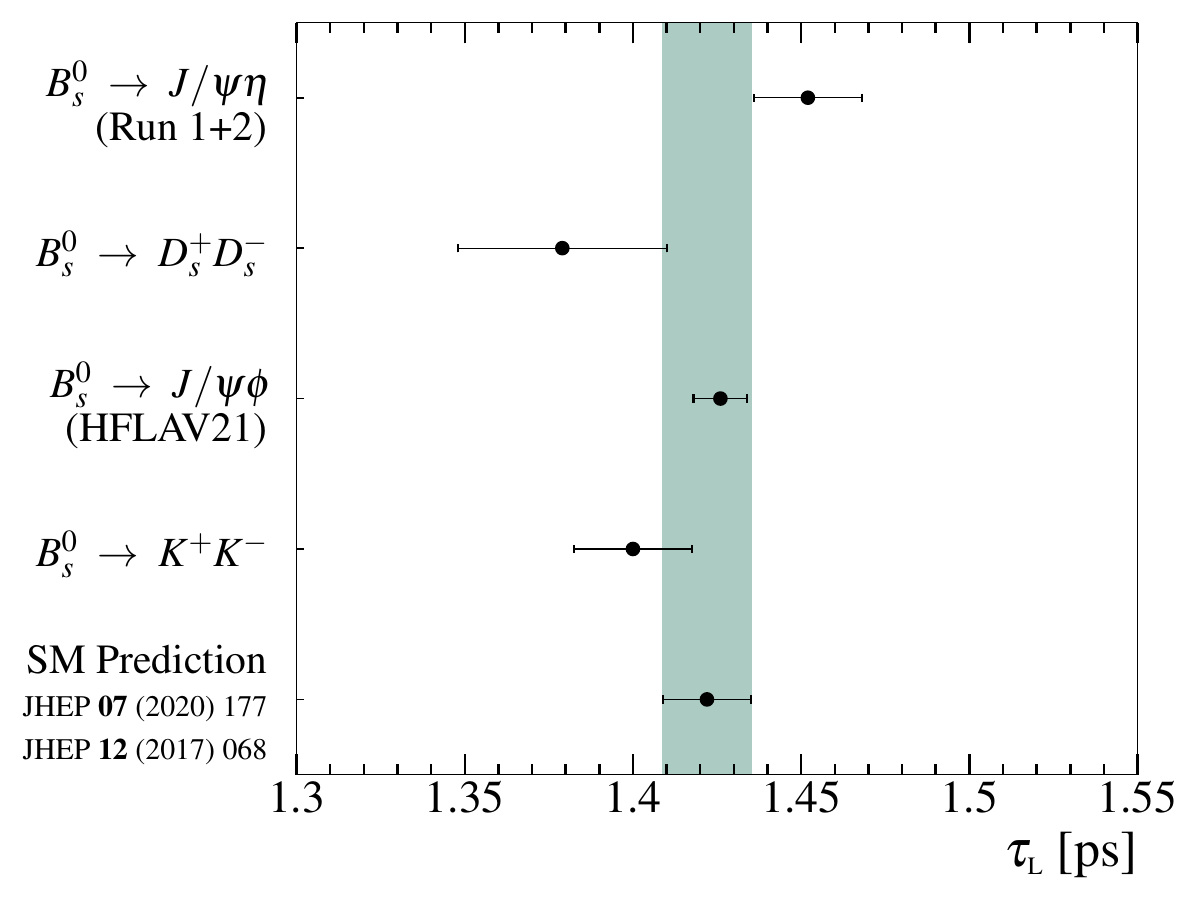}}
\caption{\small Summary of measurements of $\tau_{\rm L}$ from LHCb \cite{LHCb-PAPER-2013-060, LHCb-PAPER-2014-011} along with the HFLAV average \cite{HFLAV18} determined using the measurements of $\Gamma_s$ and $\Delta \Gamma_s$ made using the $\Bs \rightarrow J/\psi \phi$ decay mode. The SM prediction, calculated using the values in Refs.~\cite{Lenz:2019lvd,Kirk:2017juj} is shown by the grey-green band.}
\label{fig:comp}
\end{center}
\end{figure}

Using the dataset collected by LHCb during Run 2, the effective lifetime in the $B^0_s \rightarrow \jpsi\eta$ decay mode is measured to be
\begin{displaymath}
\tau_{\rm L} = 1.445 \pm 0.016~\textrm{(stat)} \pm 0.008 ~\textrm{(syst)} \ps. \nonumber
\end{displaymath}
This result agrees with the Run 1 measurement \cite{LHCb-PAPER-2016-017} within $1 \sigma$
\begin{displaymath}
\tau_{\rm L} = 1.479 \pm 0.034~\textrm{(stat)} \pm 0.011 ~\textrm{(syst)} \ps, \nonumber
\end{displaymath}
and is a factor of two more precise. The two values are combined assuming the uncertainties due to the momentum and length scales, $\Bd$ background, partially reconstructed background, mass model, time resolution and possible $\Bc$ component are fully correlated. The remaining systematic uncertainties, dominated by the simulated samples sizes, are taken to be uncorrelated. The combination gives
\begin{displaymath}
\tau_{\rm L} = 1.452 \pm 0.014 \pm 0.007 \pm 0.002 \ps, \nonumber
\end{displaymath}
where the first uncertainty is statistical, the second is the uncorrelated systematic uncertainty and the third is the correlated systematic uncertainty. This combination supersedes the previous result in the ${\mbox \Bs \rightarrow \jpsi \eta}$ mode. It is in agreement at 2$\sigma$, but more precise than, the measurement made using the $\Bs\rightarrow \Dsp \Dsm$ decay mode $\tau_{\rm L}
= 1.379 \pm 0.026 \stat \pm 0.017 \syst \ps$ \cite{LHCb-PAPER-2013-060}. A weighted average of the tree-level measurements gives $\tau_{\textrm{L}} =
1.437 \pm 0.014 \ps$. This agrees at the 1$\sigma$ level with the
Standard Model expectation of $\tau_{\rm L} = 1.422 \pm 0.013 \ps$ \cite{Lenz:2019lvd,Kirk:2017juj}. The result also agrees at the 2$\sigma$ level with the value quoted by HFLAV \cite{HFLAV18} based upon measurements of $\Gamma_s$ and $\Delta \Gamma_s$ in the $B^0_s
\rightarrow J/\psi\phi$ decay mode, $\tau_{\rm L} = 1.426 \pm 0.008 \ps$. Finally, it is also consistent with the value of $1.407 \pm 0.016 \stat \pm 0.007 \syst \ps$ found in the penguin-dominated decay $\Bs \rightarrow K^+ K^-$ \cite{LHCb-PAPER-2014-011}.  Figure~\ref{fig:comp} summarizes the measurements of $\tau_{\rm L}$ in all
these modes.

Further improvements in precision in $\tau_{\rm L}$ are expected both by considering other \CP-even $\Bs$ decays to final states containing $\eta$ or $\eta'$ mesons, the $\Bs \rightarrow \Dsp \Dsm$ dataset collected during Run 2 and from the larger dataset that will be collected by the upgraded LHCb detector.

\section*{Acknowledgements}
\noindent We express our gratitude to our colleagues in the CERN
accelerator departments for the excellent performance of the LHC. We
thank the technical and administrative staff at the LHCb
institutes.
We acknowledge support from CERN and from the national agencies:
CAPES, CNPq, FAPERJ and FINEP (Brazil);
MOST and NSFC (China);
CNRS/IN2P3 (France);
BMBF, DFG and MPG (Germany);
INFN (Italy);
NWO (Netherlands);
MNiSW and NCN (Poland);
MEN/IFA (Romania);
MICINN (Spain);
SNSF and SER (Switzerland);
NASU (Ukraine);
STFC (United Kingdom);
DOE NP and NSF (USA).
We acknowledge the computing resources that are provided by CERN, IN2P3
(France), KIT and DESY (Germany), INFN (Italy), SURF (Netherlands),
PIC (Spain), GridPP (United Kingdom),
CSCS (Switzerland), IFIN-HH (Romania), CBPF (Brazil),
Polish WLCG  (Poland) and NERSC (USA).
We are indebted to the communities behind the multiple open-source
software packages on which we depend.
Individual groups or members have received support from
ARC and ARDC (Australia);
Minciencias (Colombia);
AvH Foundation (Germany);
EPLANET, Marie Sk\l{}odowska-Curie Actions and ERC (European Union);
A*MIDEX, ANR, IPhU and Labex P2IO, and R\'{e}gion Auvergne-Rh\^{o}ne-Alpes (France);
Key Research Program of Frontier Sciences of CAS, CAS PIFI, CAS CCEPP,
Fundamental Research Funds for the Central Universities,
and Sci. \& Tech. Program of Guangzhou (China);
GVA, XuntaGal and GENCAT (Spain);
SRC (Sweden);
the Leverhulme Trust, the Royal Society
 and UKRI (United Kingdom).

\addcontentsline{toc}{section}{References}
\ifx\mcitethebibliography\mciteundefinedmacro
\PackageError{LHCb.bst}{mciteplus.sty has not been loaded}
{This bibstyle requires the use of the mciteplus package.}\fi
\providecommand{\href}[2]{#2}

\newpage

\centerline
{\large\bf LHCb collaboration}
\begin
{flushleft}
\small
R.~Aaij$^{32}$\lhcborcid{0000-0003-0533-1952},
A.S.W.~Abdelmotteleb$^{50}$\lhcborcid{0000-0001-7905-0542},
C.~Abellan~Beteta$^{44}$,
F.~Abudin{\'e}n$^{50}$\lhcborcid{0000-0002-6737-3528},
T.~Ackernley$^{54}$\lhcborcid{0000-0002-5951-3498},
B.~Adeva$^{40}$\lhcborcid{0000-0001-9756-3712},
M.~Adinolfi$^{48}$\lhcborcid{0000-0002-1326-1264},
H.~Afsharnia$^{9}$,
C.~Agapopoulou$^{13}$\lhcborcid{0000-0002-2368-0147},
C.A.~Aidala$^{76}$\lhcborcid{0000-0001-9540-4988}
% ...

{\footnotesize \it

% ...
$^{9}$Universit{\'e} Clermont Auvergne, CNRS/IN2P3, LPC, Clermont-Ferrand, France\\
% ...
$^{13}$LPNHE, Sorbonne Universit{\'e}, Paris Diderot Sorbonne Paris Cit{\'e}, CNRS/IN2P3, Paris, France\\
% ...
$^{32}$Nikhef National Institute for Subatomic Physics, Amsterdam, Netherlands\\
% ...
$^{40}$Instituto Galego de F{\'\i}sica de Altas Enerx{\'\i}as (IGFAE), Universidade de Santiago de Compostela, Santiago de Compostela, Spain\\
% ...
$^{44}$Physik-Institut, Universit{\"a}t Z{\"u}rich, Z{\"u}rich, Switzerland\\
% ...
$^{48}$H.H. Wills Physics Laboratory, University of Bristol, Bristol, United Kingdom\\
% ...
$^{50}$Department of Physics, University of Warwick, Coventry, United Kingdom\\
% ...
$^{54}$Oliver Lodge Laboratory, University of Liverpool, Liverpool, United Kingdom\\
% ...
$^{62}$Syracuse University, Syracuse, NY, United States\\
% ...
$^{76}$University of Michigan, Ann Arbor, MI, United States, associated to $^{62}$\\
\bigskip
% ...
}
\end{flushleft}


\begin{mcitethebibliography}{10}
\mciteSetBstSublistMode{n}
\mciteSetBstMaxWidthForm{subitem}{\alph{mcitesubitemcount})}
\mciteSetBstSublistLabelBeginEnd{\mcitemaxwidthsubitemform\space}
{\relax}{\relax}

\bibitem{Lenz:2019lvd}
A.~Lenz and G.~Tetlalmatzi-Xolocotzi,
  \ifthenelse{\boolean{articletitles}}{\emph{{Model-independent bounds on new
  physics effects in non-leptonic tree-level decays of B-mesons}},
  }{}\href{https://doi.org/10.1007/JHEP07(2020)177}{JHEP \textbf{07} (2020)
  177}, \href{http://arxiv.org/abs/1912.07621}{{\normalfont\ttfamily
  arXiv:1912.07621}}\relax
\mciteBstWouldAddEndPuncttrue
\mciteSetBstMidEndSepPunct{\mcitedefaultmidpunct}
{\mcitedefaultendpunct}{\mcitedefaultseppunct}\relax
\EndOfBibitem
\bibitem{Fleischer:2011cw}
R.~Fleischer and R.~Knegjens,
  \ifthenelse{\boolean{articletitles}}{\emph{{Effective lifetimes of $B_s$
  decays and their constraints on the $B_s^0$-$\bar B_s^0$ mixing parameters}},
  }{}\href{https://doi.org/10.1140/epjc/s10052-011-1789-9}{Eur.\ Phys.\ J.\
  \textbf{C71} (2011) 1789},
  \href{http://arxiv.org/abs/1109.5115}{{\normalfont\ttfamily
  arXiv:1109.5115}}\relax
\mciteBstWouldAddEndPuncttrue
\mciteSetBstMidEndSepPunct{\mcitedefaultmidpunct}
{\mcitedefaultendpunct}{\mcitedefaultseppunct}\relax
\EndOfBibitem
\bibitem{Fleischer:2011ib}
R.~Fleischer, R.~Knegjens, and G.~Ricciardi,
  \ifthenelse{\boolean{articletitles}}{\emph{{Exploring CP violation and
  $\eta$-$\eta'$ mixing with the $B^0_{s,d} \to J/\psi \eta^{(\prime)}$
  systems}}, }{}\href{https://doi.org/10.1140/epjc/s10052-011-1798-8}{Eur.\
  Phys.\ J.\  \textbf{C71} (2011) 1798},
  \href{http://arxiv.org/abs/1110.5490}{{\normalfont\ttfamily
  arXiv:1110.5490}}\relax
\mciteBstWouldAddEndPuncttrue
\mciteSetBstMidEndSepPunct{\mcitedefaultmidpunct}
{\mcitedefaultendpunct}{\mcitedefaultseppunct}\relax
\EndOfBibitem
\bibitem{Kirk:2017juj}
M.~Kirk, A.~Lenz, and T.~Rauh,
  \ifthenelse{\boolean{articletitles}}{\emph{{Dimension-six matrix elements for
  meson mixing and lifetimes from sum rules}},
  }{}\href{https://doi.org/10.1007/JHEP12(2017)068}{JHEP \textbf{12} (2017)
  068}, Erratum \href{https://doi.org/10.1007/JHEP06(2020)162}{ibid.\
  \textbf{06} (2020) 162},
  \href{http://arxiv.org/abs/1711.02100}{{\normalfont\ttfamily
  arXiv:1711.02100}}\relax
\mciteBstWouldAddEndPuncttrue
\mciteSetBstMidEndSepPunct{\mcitedefaultmidpunct}
{\mcitedefaultendpunct}{\mcitedefaultseppunct}\relax
\EndOfBibitem
\bibitem{PDG2020}
Particle Data Group, P.~A. Zyla {\em et~al.},
  \ifthenelse{\boolean{articletitles}}{\emph{{\href{http://pdg.lbl.gov/}{Review
  of particle physics}}}, }{}\href{https://doi.org/10.1093/ptep/ptaa104}{Prog.\
  Theor.\ Exp.\ Phys.\  \textbf{2020} (2020) 083C01}\relax
\mciteBstWouldAddEndPuncttrue
\mciteSetBstMidEndSepPunct{\mcitedefaultmidpunct}
{\mcitedefaultendpunct}{\mcitedefaultseppunct}\relax
\EndOfBibitem
\bibitem{LHCb-PAPER-2016-017}
LHCb collaboration, R.~Aaij {\em et~al.},
  \ifthenelse{\boolean{articletitles}}{\emph{{Measurement of the
  \mbox{\decay{\Bs}{\jpsi\eta}} lifetime}},
  }{}\href{https://doi.org/10.1016/j.physletb.2016.10.006}{Phys.\ Lett.\
  \textbf{B762} (2016) 484},
  \href{http://arxiv.org/abs/1607.06314}{{\normalfont\ttfamily
  arXiv:1607.06314}}\relax
\mciteBstWouldAddEndPuncttrue
\mciteSetBstMidEndSepPunct{\mcitedefaultmidpunct}
{\mcitedefaultendpunct}{\mcitedefaultseppunct}\relax
\EndOfBibitem
\bibitem{LHCb-PAPER-2013-060}
LHCb collaboration, R.~Aaij {\em et~al.},
  \ifthenelse{\boolean{articletitles}}{\emph{{Measurement of the
  \mbox{\decay{\Bsb}{\Dsm\Dsp}} and \mbox{\decay{\Bsb}{\Dm\Dsp}} effective
  lifetimes}}, }{}\href{https://doi.org/10.1103/PhysRevLett.112.111802}{Phys.\
  Rev.\ Lett.\  \textbf{112} (2014) 111802},
  \href{http://arxiv.org/abs/1312.1217}{{\normalfont\ttfamily
  arXiv:1312.1217}}\relax
\mciteBstWouldAddEndPuncttrue
\mciteSetBstMidEndSepPunct{\mcitedefaultmidpunct}
{\mcitedefaultendpunct}{\mcitedefaultseppunct}\relax
\EndOfBibitem
\bibitem{LHCb-PAPER-2014-011}
LHCb collaboration, R.~Aaij {\em et~al.},
  \ifthenelse{\boolean{articletitles}}{\emph{{Effective lifetime measurements
  in the \mbox{\decay{\Bs}{\Kp\Km}}, \mbox{\decay{\Bz}{\Kp\pim}} and
  \mbox{\decay{\Bs}{\pip\Km}} decays}},
  }{}\href{https://doi.org/10.1016/j.physletb.2014.07.051}{Phys.\ Lett.\
  \textbf{B736} (2014) 446},
  \href{http://arxiv.org/abs/1406.7204}{{\normalfont\ttfamily
  arXiv:1406.7204}}\relax
\mciteBstWouldAddEndPuncttrue
\mciteSetBstMidEndSepPunct{\mcitedefaultmidpunct}
{\mcitedefaultendpunct}{\mcitedefaultseppunct}\relax
\EndOfBibitem
\bibitem{LHCb-PAPER-2012-017}
LHCb collaboration, R.~Aaij {\em et~al.},
  \ifthenelse{\boolean{articletitles}}{\emph{{Measurement of the \Bsb effective
  lifetime in the $\jpsi f_0(980)$ final state}},
  }{}\href{https://doi.org/10.1103/PhysRevLett.109.152002}{Phys.\ Rev.\ Lett.\
  \textbf{109} (2012) 152002},
  \href{http://arxiv.org/abs/1207.0878}{{\normalfont\ttfamily
  arXiv:1207.0878}}\relax
\mciteBstWouldAddEndPuncttrue
\mciteSetBstMidEndSepPunct{\mcitedefaultmidpunct}
{\mcitedefaultendpunct}{\mcitedefaultseppunct}\relax
\EndOfBibitem
\bibitem{D0:2016nbv}
D0 collaboration, V.~M. Abazov {\em et~al.},
  \ifthenelse{\boolean{articletitles}}{\emph{{$B^{0}_{s}$ lifetime measurement
  in the CP-odd decay channel $B^{0}_{s} \to J/\psi f_{0}(980)$}},
  }{}\href{https://doi.org/10.1103/PhysRevD.94.012001}{Phys.\ Rev.\
  \textbf{D94} (2016) 012001},
  \href{http://arxiv.org/abs/1603.01302}{{\normalfont\ttfamily
  arXiv:1603.01302}}\relax
\mciteBstWouldAddEndPuncttrue
\mciteSetBstMidEndSepPunct{\mcitedefaultmidpunct}
{\mcitedefaultendpunct}{\mcitedefaultseppunct}\relax
\EndOfBibitem
\bibitem{CMS:2017ygm}
CMS collaboration, A.~M. Sirunyan {\em et~al.},
  \ifthenelse{\boolean{articletitles}}{\emph{{Measurement of b hadron lifetimes
  in pp collisions at $\sqrt{s} =$ 8 TeV}},
  }{}\href{https://doi.org/10.1140/epjc/s10052-018-5929-3}{Eur.\ Phys.\ J.\ C
  \textbf{78} (2018) 457}, Erratum
  \href{https://doi.org/10.1140/epjc/s10052-018-6014-7}{ibid.\   \textbf{C78}
  (2018) 561}, \href{http://arxiv.org/abs/1710.08949}{{\normalfont\ttfamily
  arXiv:1710.08949}}\relax
\mciteBstWouldAddEndPuncttrue
\mciteSetBstMidEndSepPunct{\mcitedefaultmidpunct}
{\mcitedefaultendpunct}{\mcitedefaultseppunct}\relax
\EndOfBibitem
\bibitem{LHCb-DP-2012-002}
A.~A. Alves~Jr.\ {\em et~al.},
  \ifthenelse{\boolean{articletitles}}{\emph{{Performance of the LHCb muon
  system}}, }{}\href{https://doi.org/10.1088/1748-0221/8/02/P02022}{JINST
  \textbf{8} (2013) P02022},
  \href{http://arxiv.org/abs/1211.1346}{{\normalfont\ttfamily
  arXiv:1211.1346}}\relax
\mciteBstWouldAddEndPuncttrue
\mciteSetBstMidEndSepPunct{\mcitedefaultmidpunct}
{\mcitedefaultendpunct}{\mcitedefaultseppunct}\relax
\EndOfBibitem
\bibitem{LHCb-DP-2014-002}
LHCb collaboration, R.~Aaij {\em et~al.},
  \ifthenelse{\boolean{articletitles}}{\emph{{LHCb detector performance}},
  }{}\href{https://doi.org/10.1142/S0217751X15300227}{Int.\ J.\ Mod.\ Phys.\
  \textbf{A30} (2015) 1530022},
  \href{http://arxiv.org/abs/1412.6352}{{\normalfont\ttfamily
  arXiv:1412.6352}}\relax
\mciteBstWouldAddEndPuncttrue
\mciteSetBstMidEndSepPunct{\mcitedefaultmidpunct}
{\mcitedefaultendpunct}{\mcitedefaultseppunct}\relax
\EndOfBibitem
\bibitem{LHCb-PAPER-2013-011}
LHCb collaboration, R.~Aaij {\em et~al.},
  \ifthenelse{\boolean{articletitles}}{\emph{{Precision measurement of \D meson
  mass differences}}, }{}\href{https://doi.org/10.1007/JHEP06(2013)065}{JHEP
  \textbf{06} (2013) 065},
  \href{http://arxiv.org/abs/1304.6865}{{\normalfont\ttfamily
  arXiv:1304.6865}}\relax
\mciteBstWouldAddEndPuncttrue
\mciteSetBstMidEndSepPunct{\mcitedefaultmidpunct}
{\mcitedefaultendpunct}{\mcitedefaultseppunct}\relax
\EndOfBibitem
\bibitem{LHCb-DP-2020-001}
C.~Abellan~Beteta {\em et~al.},
  \ifthenelse{\boolean{articletitles}}{\emph{{Calibration and performance of
  the LHCb calorimeters in Run 1 and 2 at the LHC}},
  }{}\href{http://arxiv.org/abs/2008.11556}{{\normalfont\ttfamily
  arXiv:2008.11556}}, {submitted to JINST}\relax
\mciteBstWouldAddEndPuncttrue
\mciteSetBstMidEndSepPunct{\mcitedefaultmidpunct}
{\mcitedefaultendpunct}{\mcitedefaultseppunct}\relax
\EndOfBibitem
\bibitem{LHCb-DP-2019-001}
R.~Aaij {\em et~al.}, \ifthenelse{\boolean{articletitles}}{\emph{{Performance
  of the LHCb trigger and full real-time reconstruction in Run 2 of the LHC}},
  }{}\href{https://doi.org/10.1088/1748-0221/14/04/P04013}{JINST \textbf{14}
  (2019) P04013}, \href{http://arxiv.org/abs/1812.10790}{{\normalfont\ttfamily
  arXiv:1812.10790}}\relax
\mciteBstWouldAddEndPuncttrue
\mciteSetBstMidEndSepPunct{\mcitedefaultmidpunct}
{\mcitedefaultendpunct}{\mcitedefaultseppunct}\relax
\EndOfBibitem
\bibitem{Borghi:2017hfp}
S.~Borghi, \ifthenelse{\boolean{articletitles}}{\emph{{Novel real-time
  alignment and calibration of the LHCb detector and its performance}},
  }{}\href{https://doi.org/10.1016/j.nima.2016.06.050}{Nucl.\ Instrum.\ Meth.\
  A \textbf{845} (2017) 560}\relax
\mciteBstWouldAddEndPuncttrue
\mciteSetBstMidEndSepPunct{\mcitedefaultmidpunct}
{\mcitedefaultendpunct}{\mcitedefaultseppunct}\relax
\EndOfBibitem
\bibitem{Sjostrand:2006za}
T.~Sj\"{o}strand, S.~Mrenna, and P.~Skands,
  \ifthenelse{\boolean{articletitles}}{\emph{{PYTHIA 6.4 physics and manual}},
  }{}\href{https://doi.org/10.1088/1126-6708/2006/05/026}{JHEP \textbf{05}
  (2006) 026}, \href{http://arxiv.org/abs/hep-ph/0603175}{{\normalfont\ttfamily
  arXiv:hep-ph/0603175}}\relax
\mciteBstWouldAddEndPuncttrue
\mciteSetBstMidEndSepPunct{\mcitedefaultmidpunct}
{\mcitedefaultendpunct}{\mcitedefaultseppunct}\relax
\EndOfBibitem
\bibitem{Sjostrand:2007gs}
T.~Sj\"{o}strand, S.~Mrenna, and P.~Skands,
  \ifthenelse{\boolean{articletitles}}{\emph{{A brief introduction to PYTHIA
  8.1}}, }{}\href{https://doi.org/10.1016/j.cpc.2008.01.036}{Comput.\ Phys.\
  Commun.\  \textbf{178} (2008) 852},
  \href{http://arxiv.org/abs/0710.3820}{{\normalfont\ttfamily
  arXiv:0710.3820}}\relax
\mciteBstWouldAddEndPuncttrue
\mciteSetBstMidEndSepPunct{\mcitedefaultmidpunct}
{\mcitedefaultendpunct}{\mcitedefaultseppunct}\relax
\EndOfBibitem
\bibitem{LHCb-PROC-2010-056}
I.~Belyaev {\em et~al.}, \ifthenelse{\boolean{articletitles}}{\emph{{Handling
  of the generation of primary events in Gauss, the LHCb simulation
  framework}}, }{}\href{https://doi.org/10.1088/1742-6596/331/3/032047}{J.\
  Phys.\ Conf.\ Ser.\  \textbf{331} (2011) 032047}\relax
\mciteBstWouldAddEndPuncttrue
\mciteSetBstMidEndSepPunct{\mcitedefaultmidpunct}
{\mcitedefaultendpunct}{\mcitedefaultseppunct}\relax
\EndOfBibitem
\bibitem{Lange:2001uf}
D.~J. Lange, \ifthenelse{\boolean{articletitles}}{\emph{{The EvtGen particle
  decay simulation package}},
  }{}\href{https://doi.org/10.1016/S0168-9002(01)00089-4}{Nucl.\ Instrum.\
  Meth.\  \textbf{A462} (2001) 152}\relax
\mciteBstWouldAddEndPuncttrue
\mciteSetBstMidEndSepPunct{\mcitedefaultmidpunct}
{\mcitedefaultendpunct}{\mcitedefaultseppunct}\relax
\EndOfBibitem
\bibitem{Golonka:2005pn}
P.~Golonka and Z.~Was, \ifthenelse{\boolean{articletitles}}{\emph{{PHOTOS Monte
  Carlo: A precision tool for QED corrections in $Z$ and $W$ decays}},
  }{}\href{https://doi.org/10.1140/epjc/s2005-02396-4}{Eur.\ Phys.\ J.\
  \textbf{C45} (2006) 97},
  \href{http://arxiv.org/abs/hep-ph/0506026}{{\normalfont\ttfamily
  arXiv:hep-ph/0506026}}\relax
\mciteBstWouldAddEndPuncttrue
\mciteSetBstMidEndSepPunct{\mcitedefaultmidpunct}
{\mcitedefaultendpunct}{\mcitedefaultseppunct}\relax
\EndOfBibitem
\bibitem{Allison:2006ve}
Geant4 collaboration, J.~Allison {\em et~al.},
  \ifthenelse{\boolean{articletitles}}{\emph{{Geant4 developments and
  applications}}, }{}\href{https://doi.org/10.1109/TNS.2006.869826}{IEEE
  Trans.\ Nucl.\ Sci.\  \textbf{53} (2006) 270}\relax
\mciteBstWouldAddEndPuncttrue
\mciteSetBstMidEndSepPunct{\mcitedefaultmidpunct}
{\mcitedefaultendpunct}{\mcitedefaultseppunct}\relax
\EndOfBibitem
\bibitem{Agostinelli:2002hh}
Geant4 collaboration, S.~Agostinelli {\em et~al.},
  \ifthenelse{\boolean{articletitles}}{\emph{{Geant4: A simulation toolkit}},
  }{}\href{https://doi.org/10.1016/S0168-9002(03)01368-8}{Nucl.\ Instrum.\
  Meth.\  \textbf{A506} (2003) 250}\relax
\mciteBstWouldAddEndPuncttrue
\mciteSetBstMidEndSepPunct{\mcitedefaultmidpunct}
{\mcitedefaultendpunct}{\mcitedefaultseppunct}\relax
\EndOfBibitem
\bibitem{LHCb-PROC-2011-006}
M.~Clemencic {\em et~al.}, \ifthenelse{\boolean{articletitles}}{\emph{{The
  \lhcb simulation application, Gauss: Design, evolution and experience}},
  }{}\href{https://doi.org/10.1088/1742-6596/331/3/032023}{J.\ Phys.\ Conf.\
  Ser.\  \textbf{331} (2011) 032023}\relax
\mciteBstWouldAddEndPuncttrue
\mciteSetBstMidEndSepPunct{\mcitedefaultmidpunct}
{\mcitedefaultendpunct}{\mcitedefaultseppunct}\relax
\EndOfBibitem
\bibitem{Cowan:2016tnm}
G.~A. Cowan, D.~C. Craik, and M.~D. Needham,
  \ifthenelse{\boolean{articletitles}}{\emph{{RapidSim: an application for the
  fast simulation of heavy-quark hadron decays}},
  }{}\href{https://doi.org/10.1016/j.cpc.2017.01.029}{Comput.\ Phys.\ Commun.\
  \textbf{214} (2017) 239},
  \href{http://arxiv.org/abs/1612.07489}{{\normalfont\ttfamily
  arXiv:1612.07489}}\relax
\mciteBstWouldAddEndPuncttrue
\mciteSetBstMidEndSepPunct{\mcitedefaultmidpunct}
{\mcitedefaultendpunct}{\mcitedefaultseppunct}\relax
\EndOfBibitem
\bibitem{Hulsbergen:2005pu}
W.~D. Hulsbergen, \ifthenelse{\boolean{articletitles}}{\emph{{Decay chain
  fitting with a Kalman filter}},
  }{}\href{https://doi.org/10.1016/j.nima.2005.06.078}{Nucl.\ Instrum.\ Meth.\
  \textbf{A552} (2005) 566},
  \href{http://arxiv.org/abs/physics/0503191}{{\normalfont\ttfamily
  arXiv:physics/0503191}}\relax
\mciteBstWouldAddEndPuncttrue
\mciteSetBstMidEndSepPunct{\mcitedefaultmidpunct}
{\mcitedefaultendpunct}{\mcitedefaultseppunct}\relax
\EndOfBibitem
\bibitem{Hocker:2007ht}
H.~Voss, A.~Hoecker, J.~Stelzer, and F.~Tegenfeldt,
  \ifthenelse{\boolean{articletitles}}{\emph{{TMVA - Toolkit for Multivariate
  Data Analysis with ROOT}}, }{}\href{https://doi.org/10.22323/1.050.0040}{PoS
  \textbf{ACAT} (2007) 040}\relax
\mciteBstWouldAddEndPuncttrue
\mciteSetBstMidEndSepPunct{\mcitedefaultmidpunct}
{\mcitedefaultendpunct}{\mcitedefaultseppunct}\relax
\EndOfBibitem
\bibitem{Pivk:2004ty}
M.~Pivk and F.~R. Le~Diberder,
  \ifthenelse{\boolean{articletitles}}{\emph{{sPlot: A statistical tool to
  unfold data distributions}},
  }{}\href{https://doi.org/10.1016/j.nima.2005.08.106}{Nucl.\ Instrum.\ Meth.\
  \textbf{A555} (2005) 356},
  \href{http://arxiv.org/abs/physics/0402083}{{\normalfont\ttfamily
  arXiv:physics/0402083}}\relax
\mciteBstWouldAddEndPuncttrue
\mciteSetBstMidEndSepPunct{\mcitedefaultmidpunct}
{\mcitedefaultendpunct}{\mcitedefaultseppunct}\relax
\EndOfBibitem
\bibitem{LHCb-PAPER-2013-065}
LHCb collaboration, R.~Aaij {\em et~al.},
  \ifthenelse{\boolean{articletitles}}{\emph{{Measurements of the \Bp, \Bz, \Bs
  meson and \Lb baryon lifetimes}},
  }{}\href{https://doi.org/10.1007/JHEP04(2014)114}{JHEP \textbf{04} (2014)
  114}, \href{http://arxiv.org/abs/1402.2554}{{\normalfont\ttfamily
  arXiv:1402.2554}}\relax
\mciteBstWouldAddEndPuncttrue
\mciteSetBstMidEndSepPunct{\mcitedefaultmidpunct}
{\mcitedefaultendpunct}{\mcitedefaultseppunct}\relax
\EndOfBibitem
\bibitem{Skwarnicki:1986xj}
T.~Skwarnicki, {\em {A study of the radiative cascade transitions between the
  Upsilon-prime and Upsilon resonances}}, PhD thesis, Institute of Nuclear
  Physics, Krakow, 1986,
  {\href{http://inspirehep.net/record/230779/}{DESY-F31-86-02}}\relax
\mciteBstWouldAddEndPuncttrue
\mciteSetBstMidEndSepPunct{\mcitedefaultmidpunct}
{\mcitedefaultendpunct}{\mcitedefaultseppunct}\relax
\EndOfBibitem
\bibitem{Bukin}
BABAR collaboration, J.~P. Lees {\em et~al.},
  \ifthenelse{\boolean{articletitles}}{\emph{{Branching fraction measurements
  of the color-suppressed decays $\overline{B}^0 \to D^{(*)0} \pi^0$, $D^{(*)0}
  \eta$, $D^{(*)0} \omega$, and $D^{(*)0} \eta^\prime$ and measurement of the
  polarization in the decay $\overline{B}^0 \to D^{*0} \omega$}},
  }{}\href{https://doi.org/10.1103/PhysRevD.84.112007}{Phys.\ Rev.\
  \textbf{D84} (2011) 112007}, Erratum
  \href{https://doi.org/10.1103/PhysRevD.87.039901}{ibid.\   \textbf{D87}
  (2013) 039901}, \href{http://arxiv.org/abs/1107.5751}{{\normalfont\ttfamily
  arXiv:1107.5751}}\relax
\mciteBstWouldAddEndPuncttrue
\mciteSetBstMidEndSepPunct{\mcitedefaultmidpunct}
{\mcitedefaultendpunct}{\mcitedefaultseppunct}\relax
\EndOfBibitem
\bibitem{jbes.2009.06161}
C.~Hansen, J.~B. McDonald, and W.~K. Newey,
  \ifthenelse{\boolean{articletitles}}{\emph{Instrumental variables estimation
  with flexible distributions},
  }{}\href{https://doi.org/10.1198/jbes.2009.06161}{Journal of Business \&
  Economic Statistics \textbf{28} (2010) 13}\relax
\mciteBstWouldAddEndPuncttrue
\mciteSetBstMidEndSepPunct{\mcitedefaultmidpunct}
{\mcitedefaultendpunct}{\mcitedefaultseppunct}\relax
\EndOfBibitem
\bibitem{mnsc.44.12.1650}
P.~Theodossiou, \ifthenelse{\boolean{articletitles}}{\emph{Financial data and
  the skewed generalized t distribution},
  }{}\href{https://doi.org/10.1287/mnsc.44.12.1650}{Management Science
  \textbf{44} (1998) 1650}\relax
\mciteBstWouldAddEndPuncttrue
\mciteSetBstMidEndSepPunct{\mcitedefaultmidpunct}
{\mcitedefaultendpunct}{\mcitedefaultseppunct}\relax
\EndOfBibitem
\bibitem{HFLAV18}
Heavy Flavor Averaging Group, Y.~Amhis {\em et~al.},
  \ifthenelse{\boolean{articletitles}}{\emph{{Averages of $b$-hadron,
  $c$-hadron, and $\tau$-lepton properties as of 2018}},
  }{}\href{https://doi.org/10.1140/epjc/s10052-020-8156-7}{Eur.\ Phys.\ J.\
  \textbf{C81} (2021) 226},
  \href{http://arxiv.org/abs/1909.12524}{{\normalfont\ttfamily
  arXiv:1909.12524}}, {updated results and plots available at
  \href{https://hflav.web.cern.ch}{{\texttt{https://hflav.web.cern.ch}}}}\relax
\mciteBstWouldAddEndPuncttrue
\mciteSetBstMidEndSepPunct{\mcitedefaultmidpunct}
{\mcitedefaultendpunct}{\mcitedefaultseppunct}\relax
\EndOfBibitem
\bibitem{Dauncey:2014xga}
P.~D. Dauncey, M.~Kenzie, N.~Wardle, and G.~J. Davies,
  \ifthenelse{\boolean{articletitles}}{\emph{{Handling uncertainties in
  background shapes: the discrete profiling method}},
  }{}\href{https://doi.org/10.1088/1748-0221/10/04/P04015}{JINST \textbf{10}
  (2015) P04015}, \href{http://arxiv.org/abs/1408.6865}{{\normalfont\ttfamily
  arXiv:1408.6865}}\relax
\mciteBstWouldAddEndPuncttrue
\mciteSetBstMidEndSepPunct{\mcitedefaultmidpunct}
{\mcitedefaultendpunct}{\mcitedefaultseppunct}\relax
\EndOfBibitem
\bibitem{LHCb-PAPER-2020-046}
LHCb collaboration, R.~Aaij {\em et~al.},
  \ifthenelse{\boolean{articletitles}}{\emph{{Precise measurement of the
  $f_s/f_d$ ratio of fragmentation fractions and of $B^0_s$ decay branching
  fractions}}, }{}\href{https://doi.org/10.1103/PhysRevD.104.032005}{Phys.\
  Rev.\  \textbf{D104} (2021) 032005},
  \href{http://arxiv.org/abs/2103.06810}{{\normalfont\ttfamily
  arXiv:2103.06810}}\relax
\mciteBstWouldAddEndPuncttrue
\mciteSetBstMidEndSepPunct{\mcitedefaultmidpunct}
{\mcitedefaultendpunct}{\mcitedefaultseppunct}\relax
\EndOfBibitem
\bibitem{LHCb-PAPER-2019-013}
LHCb collaboration, R.~Aaij {\em et~al.},
  \ifthenelse{\boolean{articletitles}}{\emph{{Updated measurement of
  time-dependent \CP-violating observables in \mbox{\decay{\Bs}{\jpsi \Kp\Km}}
  decays}}, }{}\href{https://doi.org/10.1140/epjc/s10052-019-7159-8}{Eur.\
  Phys.\ J.\  \textbf{C79} (2019) 706}, Erratum
  \href{https://doi.org/10.1140/epjc/s10052-020-7875-0}{ibid.\   \textbf{C80}
  (2020) 601}, \href{http://arxiv.org/abs/1906.08356}{{\normalfont\ttfamily
  arXiv:1906.08356}}\relax
\mciteBstWouldAddEndPuncttrue
\mciteSetBstMidEndSepPunct{\mcitedefaultmidpunct}
{\mcitedefaultendpunct}{\mcitedefaultseppunct}\relax
\EndOfBibitem
\end{mcitethebibliography}
\end{document}